\documentclass[a4paper,11pt]{article}
\pdfoutput=1
\usepackage{jheppub}
\usepackage[dvipsnames]{xcolor}
\usepackage[utf8]{inputenc}
\usepackage[T5]{fontenc}
\usepackage[english]{babel}
\usepackage{mathtools}
\usepackage{amsfonts}
\usepackage{amsthm}
\usepackage{physics}
\usepackage{comment}
\usepackage{float}
\usepackage{subcaption}
\usepackage{tikz}
\usepackage{tkz-euclide}
\usetikzlibrary{decorations.markings}
\usetikzlibrary{patterns}
\usepackage{tikz-feynman}
\usepackage{simpler-wick}
\usepackage{booktabs} 
\usepackage{hyperref}
\usepackage{mathrsfs}
\hypersetup{urlcolor=magenta, citecolor=Maroon, linkcolor=NavyBlue}
\usepackage{url}

\allowdisplaybreaks
\renewcommand{\a}{\alpha}

\newcommand{\g}{\gamma}
\renewcommand{\d}{\delta}

\renewcommand{\th}{\theta}
\renewcommand{\l}{\lambda}
\newcommand{\s}{\sigma}
\renewcommand{\t}{\tau}
\newcommand{\D}{\Delta}

\renewcommand{\mathbf}[1]{\textbf{#1}} 
\newcommand{\DeclareAutoPairedDelimiter}[3]{%
  \expandafter\DeclarePairedDelimiter\csname Auto\string#1\endcsname{#2}{#3}%
  \begingroup\edef\x{\endgroup
    \noexpand\DeclareRobustCommand{\noexpand#1}{%
      \expandafter\noexpand\csname Auto\string#1\endcsname*}}%
  \x}
\DeclareAutoPairedDelimiter{\p}{(}{)}

\usepackage{bbm}
\usepackage{bm} 
\renewcommand{\it}[1]{\textit{#1}} 

\usepackage{cancel} 

\renewcommand{\tilde}[1]{\widetilde{#1}}
\DeclarePairedDelimiterX{\avg}[1]{\langle}{\rangle}{#1} 

\theoremstyle{plain}

\newcommand{\bp}{\bar\partial}
\newcommand{\RR}{\mathbb R}
\newcommand{\CC}{\mathbb C}
\newcommand{\ZZ}{\mathbb Z}

\newcommand{\im}{\operatorname{Im}}
\newcommand{\re}{\operatorname{Re}}

\author{Ziyi Li}
\affiliation{Department of Physics, University of California at Davis, CA 95616, U.S.A.}
\emailAdd{zzyli@ucdavis.edu}

\title{Black Strings and Black Holes: Worldsheet Spectra and Near-Extremal Limits}

\abstract{We study aspects of the Euclidean partition function of the $\frac{SL(2,\mathbb R)\times U(1)}{U(1)}$ gauged WZW model. The target space geometry is either a smooth horizonless geometry with a linear-dilaton asymptotic region, or the Horne--Horowitz black string, depending on which Euclidean circle we continue as the time-direction. We first study the worldsheet spectrum and obtain the character decomposition via the holonomy integrals. Next, we obtain the spacetime spectrum of the horizonless geometry by directly performing the unfolded modular integrals, and recover the Maldacena-Ooguri-Son spectrum in the ungauged limit. In the chirally gauged model, the dimensionally reduced target space is instead a two-dimensional charged black hole. We study the near extremal regime of the black hole and black string. For the two dimensional charged black hole, we identify two neutral towers of states that produce the expected $\frac{3}{2}\log T$ of the AdS$_2$ throat from the worldsheet discrete series representation. The discrete towers appear only below the temperature at which the winding sector loses absolute convergence. For the black string, we propose two towers of spatially winding strings in the discrete series that yield a $\log T^2$ contribution to the full amplitude in the near-extremal limit. We comment on the contribution of these states in the near-extremal limit of the full partition function.}
\newpage

\begin{document} 
\maketitle
\flushbottom

\section{Introduction}
String theory in low spacetime dimensions has proven to be a useful arena for exploring various aspects of quantum gravity. Certain models are exactly solvable while retaining important features of their higher dimensional counterparts, including nontrivial black hole geometries. Black hole solutions can also be constructed from the low-energy effective string actions, which formed the basis of much early work \cite{Gibbons:1985ac,Gibbons:1987ps,Garfinkle:1990qj,Horowitz:1991cd}. However, it is generally difficult to obtain an exact worldsheet description of a given black hole background. Low dimensional string theory provides a unique window into string theory on strongly curved black hole backgrounds and their worldsheet descriptions. Witten first realized that the coset $SL(2,\mathbb R)/U(1)$ gauged WZW model can be interpreted as having a 2D black hole target space \cite{Witten:1991yr}. Depending on whether we gauge a compact or non-compact $U(1)$ subgroup, we get either a Euclidean cigar or a Lorentzian black hole, respectively. The same procedure can be generalized by starting with a four dimensional group manifold $SL(2,\mathbb R)\times \mathbb R$, and gauging the appropriate $U(1)$ subgroup and a translation in $\mathbb R$ to obtain a three-dimensional black string solution \cite{Horne:1991gn}. When the gauged action is chirally charged under $U(1)$ ($\mathbb R$) group, the dimensionally reduced theory is instead a charged two-dimensional black hole solution \cite{Ishibashi:1991wh}. This geometry can also be obtained from the low energy effective action of two-dimensional string theory containing gauge fields \cite{McGuigan:1991qp}. More interestingly, the three dimensional black string can be shown to be equivalent to the three dimensional BTZ black hole \cite{Banados:1992wn,Banados:1992gq} under $T$-duality \cite{Horowitz:1993jc}, demonstrating that a slight modification of the BTZ black hole with non-vanishing three form field strength $H_{\mu\nu\g}\propto\epsilon_{\mu \nu \g}$ is an exact solution to string theory \cite{Horowitz:1993jc,Kaloper:1993kj}.

Despite the success in finding an exact conformal field theory describing black holes, the non-compact nature of $SL(2,\mathbb R)$ makes the studies of its current algebra rather subtle \cite{Gawedzki:1991yu,Dijkgraaf:1991ba,Giveon:1998ns}. It was not until the seminal studies of Maldacena, Ooguri, and Son \cite{Maldacena:2000hw,Maldacena:2000kv,Maldacena:2001km} that made clear the importance of spectrally flowed string states that rendered the quantum theory consistent. Their study has since opened a new path into the study of integrable systems in AdS$_3$ string theory with NS-NS flux and directly paved the way to the later advances in the studies of tensionless string theories \cite{Eberhardt:2018ouy,Eberhardt:2019ywk} (see \cite{Kovensky:2026usc} for a review and references therein). Their studies also allowed the investigation into the one-loop effects of the $SL(2,\mathbb R)/U(1)$ two-dimensional black hole conformal field theory, where the decomposition of the worldsheet partition function into characters was performed, finding the expected continuous and discrete representations \cite{Hanany:2002ev}. 

It is therefore natural to study the richer coset theory that describes the three dimensional black string and charged two dimensional black hole next. The deformed partition function was studied in \cite{Israel:2003ry}, with more recent thermal partition function developments in \cite{Chakraborty:2024mls,Toni:2026brc}, with \cite{Toni:2026brc} having a focus on the two-dimensional black hole and its near extremal properties. In this paper, we will focus on computing the Euclidean partition function of the $\p{SL(2,\mathbb R)\times U(1)}/U(1)$ gauged WZW theory, and its worldsheet and target space spectrum. Other than the aforementioned connection to the BTZ black holes, the three dimensional black string has an interesting extremal limit. In the worldsheet description, the extremal limit of the black string leaves the $SL(2,\mathbb R)$ sector ungauged, yielding a locally AdS$_3$ target space geometry \cite{Horne:1991gn}. It is thus interesting to ask whether the partition function of Maldacena--Ooguri--Son and its spacetime spectrum can be recovered from performing the gauged Euclidean path integral. Indeed, we will show that this can be achieved, and the connection becomes more transparent if we choose the compact $U(1)$ circle as our Lorentzian time direction, which yields a smooth horizonless geometry with a linear dilaton asymptotic. This background belongs to the exactly marginal $J_3\bar J_3$ deformation of the $SL(2,\mathbb R)$ WZW model studied in \cite{Israel:2003ry,Chakraborty:2024mls}, and the extremal limit of the black string corresponds to the undeformed limit of this theory. The deformed partition function was studied in \cite{Israel:2003ry}, where the worldsheet spectrum was extracted to first order in the deformation.
The authors of \cite{Chakraborty:2024mls} reached the deformed string spectrum by imposing the gauge and Virasoro constraints and obtained a spectral expression for the amplitude, while leaving the direct evaluation of the modular integral as an open check. Here we perform the holonomy integrals directly, and obtain the worldsheet character decomposition at finite deformation, as well as the complete worldsheet spectrum for the chirally gauged model relevant for the two-dimensional black hole. For the horizonless geometry, we obtain the spacetime spectrum by unfolding the winding sum and directly evaluating the holonomy and modular integrals. We obtain a string trace whose energy is consistent with the deformed mass-shell conditions. In the undeformed limit we then recover the AdS$_3$ string spectrum. 

Another aspect of our analysis focuses on the near extremal limit of the partition function. Near-extremal behaviors of black holes have been central in our understanding of black hole quantum mechanics. One longstanding puzzle is the non-vanishing classical entropy at zero temperature. For non-supersymmetric black holes, it is unclear what symmetry might protect such a large degeneracy, or how a near-extremal black hole could Hawking radiate \cite{Preskill:1991tb}. These puzzles have been recently clarified using techniques of gravitational path integral \cite{Ghosh:2019rcj,Iliesiu:2020qvm,Heydeman:2020hhw}, where one finds that the near-extremal dynamics of these black holes are governed by the AdS$_2$ throat with an enhanced $SL(2,\mathbb R)$ symmetry which ought to be treated quantum mechanically \cite{Sen:2012cj,Sen:2012kpz,Maldacena:2016upp,Iliesiu:2022onk}. Integrating over these nearly-zero modes of the Schwarzian action yields a term behaving as $\log T^{3/2}$ \cite{Stanford:2017thb}, which ends up removing the large density of states for various non-supersymmetric black holes \cite{Kapec:2023ruw,Rakic:2023vhv,Maulik:2024dwq,Kapec:2024zdj,Maulik:2025hax}. Schematically, the near-extremal partition function looks like: 
\begin{align}
    \log Z(T)\simeq S_0+2\pi^2C T +\frac{3}{2}\log \p{CT}
\end{align}
where $S_0$ is the extremal black hole entropy and $C$ is a constant that controls the linear $T$ corrections. Given that the $\p{SL(2,\mathbb R)\times U(1)}/U(1)$ gauged WZW model can be used to describe both a charged three-dimensional black string and a two dimensional charged black hole, it would be interesting to study their near-extremal behavior from the perspective of the worldsheet. As discussed above, the three-dimensional black string has a locally AdS$_3$ target space geometry, and when compactified, it exhibits an AdS$_2$ Schwarzian throat \cite{Giveon:2004zz}, where one expects a low-temperature correction to the non-vanishing entropy. The semi-classical studies as well as the worldsheet quantum correction were carried out for the two dimensional charged black hole in \cite{Toni:2026brc}, where by mapping the results of \cite{Iliesiu:2020qvm} to the two-dimensional black hole, the semi-classical analysis shows the expected $\log T^{3/2}$ correction. On the other hand, \cite{Toni:2026brc} showed that this term was not recovered in the short-string contribution extrapolated from the partition function, which is under control for $T\gtrsim 1/(4\pi\sqrt{k})$. This threshold was then associated with
a divergent long-string contribution and was speculated to be interpreted as the black hole/string transition \cite{Susskind:1993ws,Horowitz:1996nw,Horowitz:1997jc,Giveon:2005jv,Chen:2021dsw}. 

Here, we revisit this problem of near-extremal behavior of the spectrum, and identify two towers of neutral discrete states whose weights approach one at extremality. After performing the modular integral,  we find that their combined contribution exhibits the expected Schwarzian $\log T^{3/2}$. Interestingly, these towers only enter the discrete window below $T_*=1/(4\pi \sqrt{k})$, at which the full amplitude stops being convergent. Nonetheless, the combined contribution of the discrete poles together with their accompanying continuum varies smoothly through $T_*$ and retains the $\log T^{3/2}$ in the extremal limit. We provide evidence that these states can be viewed as the spacetime Virasoro modes of the AdS$_2$ throat, and discuss various subtleties regarding the full amplitude that must be resolved before we can establish this as a universal correction to the entropy. Applying the same worldsheet analysis to the three-dimensional black string at fixed asymptotic radius $R$ in the near-extremal limit, we find two towers of spatially wound strings that yield a $\log T^2$ contribution to the one-loop amplitude.

\paragraph{Outline} The paper is organized as follows. In section \ref{Section2}, we review aspects of the $(SL(2,\mathbb R)\times U(1))/U(1)$ gauged WZW model and its target space geometries. In section \ref{section3}, we construct the worldsheet partition function and explain our holonomy conventions. We study the worldsheet spectrum and its character decomposition in section \ref{section4}, and recover the expected continuous and discrete series representations. We then move on to study the spacetime spectrum for the horizonless geometry in section \ref{section5}, where in section \ref{section5.1}, we construct the spacetime single string excitation trace using the worldsheet spectrum, and then reproduce it directly from evaluating the holonomy and unfolded modular integral in section \ref{section5.2}. We also find the ungauged limit of our spectrum reproduces the MOS analysis, along with a subtlety regarding the continuous density in section \ref{section5.3}. The details of the integral computation are given in Appendix \ref{intappendix}. Finally, in section \ref{section6}, we first construct the partition function of the two dimensional charged black hole obtained by asymmetrical gauging. Then, we identify two towers of neutral states in the discrete representation that become marginal in the extremal limit, and show that after performing the modular integral and relevant sums, they produce the expected $\log T^{3/2}$ correction. Using the same mechanism, we identify two towers of spatially wound string states that contribute a $2\log T$ to the black string amplitude. We discuss certain subtleties that must be resolved before we can establish these contributions as the universal correction to the entropy. 

\section{Worldsheet theory and its target space}
\label{Section2}

In this section, we review aspects of the conformal field theory construction that gives a three-dimensional black string target spacetime. Since we are interested in computing the Euclidean partition function, we start with a parent group $G=\mathbb H^+_3\times U(1)$ with parametrization of $H^+_3$ given by: 
\begin{equation}
 g=\begin{pmatrix}
 e^{i\widetilde\theta}\cosh r,&e^{-i\theta}\sinh r\\
 e^{i\theta}\sinh r,&e^{-i\widetilde\theta}\cosh r
 \end{pmatrix},\qquad
  \widetilde\theta\in i\RR,\quad
 \theta\sim\theta+2\pi.
\end{equation}
We compactify the additional boson on a circle of radius $R$: $Y_0\sim Y_0+2\pi R$. The action of the ungauged WZW model is given by: 
\begin{equation}
\begin{aligned}
S_{\text{ungauged}}[g,Y_0]
=\frac{k}{\pi}\int_\Sigma d^2z\,\Big[
&\partial r\,\bar\partial r +\sinh^2r\,\partial\theta\,\bar\partial\theta
-\cosh^2r\,\partial\widetilde\theta\, \bar\partial\widetilde\theta \\
&-\sinh^2r\left( \partial\widetilde\theta\,\bar\partial\theta -\partial\theta\,\bar\partial\widetilde\theta \right) +\frac{1}{k}\partial Y_0\,\bar\partial Y_0 \Big]
\end{aligned}
\end{equation}
We gauge the one-dimensional subgroup $H$ of the symmetry group of $G$ with action $g\mapsto h_\epsilon g h_\epsilon$. We choose the non-compact subgroup $h_\epsilon=\operatorname{diag}(e^{\epsilon/2},e^{-\epsilon/2})$ so that our target space, after Wick rotating the $\theta$ coordinate, is the Horne--Horowitz Lorentzian black string \cite{Horne:1991gn}. This global symmetry can be made local by making $\epsilon$ depend on the worldsheet coordinates, with transformations: 
\begin{align}
    \delta\widetilde\theta=-i\epsilon,\qquad
 \delta Y_0=c\epsilon,\qquad
 \delta A_z=\partial\epsilon,\qquad
 \delta A_{\bar z}=\bp\epsilon.
\end{align}
where we have introduced the gauge field $A_{\bar z}=A_z^*$, and the transformation of the compact boson is simply a translation. Now, the gauged action becomes: 
\begin{align}
\label{gaugedwzw}
     S_\text{gauged}=\frac{k}{\pi}\int\dd^2z\,\Big[&\partial r\bp r+\sinh^2r\,\partial\theta\bp\theta-\cosh^2r\,D_z\widetilde\theta D_{\bar z}\widetilde\theta\notag\\
     &-\sinh^2r\big(D_z\widetilde\theta\bp\theta -D_{\bar z}\widetilde\theta\partial\theta\big)+\frac1kD_zY_0D_{\bar z}Y_0\Big]
\end{align}
where we have defined $D_z\widetilde\theta=\partial\widetilde\theta+iA_z$ and
$D_zY_0=\partial Y_0-cA_z$. It is also useful to define the gauge invariant coordinate $X=Y_0-ic\widetilde\theta$, with period $2\pi R$. In order to obtain a target space geometry, we need to classically integrate out the gauge field, and then pick a gauge to read off the spacetime metric. 

\subsection{Horne-Horowitz black string}
After classically integrating out the gauge field, we fix the gauge by setting $\widetilde \theta=0$. The action is now: 
\begin{align}
    S_{\mathrm{gauged}}=\frac{k}{\pi}\int_\Sigma\dd^2z\,\Big[\partial r\,\bp r+\frac{K\sinh^2r}{D(r)}\, \partial\theta\,\bp\theta+\frac{\cosh^2r}{D(r)}\partial X\,\bp X+\frac{ic\sinh^2r}{D(r)} \big(\partial\theta\,\bp X-\partial X\,\bp\theta\big)\Big]
\end{align}
where we defined $K=k+c^2$, and $D(r)=k\cosh^2r+c^2$. The metric and $B$ field can be read off as: 
\begin{align}
\label{metric1}
    \dd s^2 =k\dd r^2+\frac{kK\sinh^2r}{D(r)}\dd\theta^2+\frac{k\cosh^2r}{D(r)}\dd X^2, \qquad B=\frac{ikc\sinh^2r}{D(r)}\dd\theta\wedge\dd X
\end{align}
The dilaton field can be solved from the classical three-dimensional low-energy string action and is given by \cite{Witten:1991yr,Horne:1991gn}: 
\begin{align}
    \Phi=\Phi_0-\frac12\log\frac{D(r)}k
\end{align}
This geometry describes the Horne--Horowitz black string if we continue $\theta=it_{\text{HH}}/\sqrt{K}$ as our Lorentzian time, and can be put into their form by the following coordinate transformation: 
\begin{equation}
 \frac{\mathcal R}{M}=\frac{D(r)}K,\qquad
 \frac QM=\frac c{\sqrt K},\qquad
 \theta=\frac{i t_{\mathrm{HH}}}{\sqrt K}.
 \label{HHmap}
\end{equation}
where the metric and $B$-field now take the form: 
\begin{align}
 \dd s^2={}&-
 \left(1-\frac M{\mathcal R}\right)\dd t_{\mathrm{HH}}^2
 +\left(1-\frac{Q^2}{M\mathcal R}\right)\dd X^2
 +\frac{k\,\dd\mathcal R^2}
 {4(\mathcal R-M)(\mathcal R-Q^2/M)},\notag\\
 B={}&-\frac QM
 \left(1-\frac M{\mathcal R}\right)
 \dd t_{\mathrm{HH}}\wedge\dd X.
\end{align}
The event horizon of the black string is located at $\mathcal{R}=M$, and $Q$ describes the charge of the black string. Note that when $c=0$ the charge $Q$ vanishes, and we simply obtain Witten's two-dimensional black hole times the compact circle. The inverse temperature of the black string is a function of $k,c$, and is given by $\beta_{HH}=2\pi \sqrt{K}$. The extremal limit of the black string is $Q\to M$, which corresponds to the $c \to \infty$ limit as is evident from the second equation of (\ref{HHmap}). Interestingly, from the perspective of the worldsheet, taking such a limit corresponds to not gauging the $SL(2,\mathbb R)$ sector, leaving us with a locally AdS$_3$ target space geometry. This can be more directly seen from the metric (\ref{metric1}). First we rescale $X=cx$ and its radius $R=c\ell$, and taking the limit $c \to \infty$ gives: 
\begin{align}
    ds^2=k\p{dr^2+\sinh^2r d\theta^2+\cosh ^2r\hspace{1mm}dx^2}
\end{align}
Note that this is thermal AdS$_3$ with curvature radius $\sqrt{k}$ and period $x\sim x+2\pi \ell$. The inverse temperature is given by the parameter $b=2\pi \ell$. This also explains why it is more natural to choose the $U(1)$ circle as the time-direction if we want to recover the spacetime spectrum of Maldacena, Ooguri, and Son \cite{Maldacena:2000hw,Maldacena:2000kv}. Next, we move on to construct the worldsheet partition function. 

\section{Worldsheet partition function}
\label{section3}
The partition function of the two-dimensional cigar was constructed by Hanany, Prezas and Troost \cite{Hanany:2002ev}, where the authors recovered the expected worldsheet characters for the discrete and continuous representations. The study of the $(SL(2,\mathbb R)\times U(1))/U(1)$ conformal field theory was initiated in \cite{Israel:2003ry}, and more recently in \cite{Chakraborty:2024mls,Toni:2026brc}, with Toni emphasizing the near-extremal effects of the two-dimensional charged black holes\footnote{Note that the construction of \cite{Hanany:2002ev} leaves the black hole temperature implicit, while \cite{Toni:2026brc} kept inverse temperature $\beta$ explicit and later identified it with the inverse black hole temperature. The two constructions match when one picks $\beta=2\pi i$ in the uncharged limit of \cite{Toni:2026brc}.}. The boson of the two-dimensional black hole is charged chirally under the $U(1)$, where one has $\bar c \neq c=0$, while for black strings one has $c=\bar c\neq0$. Here, we shall follow the constructions of Hanany et al. \cite{Hanany:2002ev} and their holonomy conventions for the gauge field. 

Let's write $z=\s_1+\t\s_2$. The area of the worldsheet torus is simply given by $4\pi^2 \t_2$. The gauge field can be Hodge decomposed as \cite{Hanany:2002ev}:
\begin{equation}
 A_z=\partial f+i\partial\widehat h-\frac{\bar u}{2{\tau_2}},\qquad
 A_{\bar z}=\bp f-i\bp\widehat h-\frac u{2{\tau_2}},\qquad
 u=u_1\tau-u_2.
 \label{tor:hodge}
\end{equation}
Note that the holonomies satisfy $u_{1,2}\in \mathbb R$, and $f,\hspace{1mm}\widehat h$ are real periodic fields. The integration over $f$ is simply the volume of the local gauge group. It is also useful to define the imaginary and real fields: 
\begin{align}
    p=-if+iL_+, \qquad h=\widehat h+iL_-
\end{align}
Since the $X$-direction is compact, strings can wind around it. To describe the winding and holonomy contributions, we introduce: 
\begin{equation}
 W=\omega\tau-m_X,\qquad
 F=\frac{iR}{2{\tau_2}}(\bar Wz-W\bar z),\qquad
 L_\pm=\frac{\bar uz\pm u\bar z}{4{\tau_2}}.
 \label{tor:harmonics}
\end{equation}
To perform the path integral, it is useful to transform the above $\mathbb H^+_3$ parametrization into the $SL(2,\mathbb C)/SU(2)$ coset model with the following coordinate transformation \cite{Gawedzki:1991yu,Maldacena:2000kv,Hanany:2002ev}:
\begin{align}
\phi=i\widetilde\vartheta-\log\cosh r, \quad V=\sinh r\,e^{i\vartheta}, \quad\bar V=\sinh r\,e^{-i\vartheta}.
\end{align}
where 
\begin{align}
    \vartheta=\theta+h, \quad \widetilde\vartheta =\widetilde\theta-p, \quad Y&=Y_0-icp
\end{align}
Substituting these coordinates back into (\ref{gaugedwzw}) yields: 
\begin{align}
    S_{\text{gauged}}=\frac{k}{\pi}\int\dd^2z\,\Big[&
 \partial\phi\bp\phi+
 (\partial\bar V+\bar V\partial\phi)
 (\bp V+V\bp\phi)+\partial h\bp h\notag\\
 &+\frac1k(\partial Y-ic\partial h)(\bp Y+ic\bp h)
 +i(\partial\phi\bp h-\partial h\bp\phi)\Big]
\end{align}
Note that all the above fields have non-trivial holonomies, and in order to integrate over these fields, we will decompose them into a periodic part and a holonomy part \cite{Hanany:2002ev}: 
\begin{align}
\label{holonomies}
     \phi&=\widehat\phi+L_+,\quad
 V=e^{-L_-}\widehat V, \quad \bar V =e^{L_-}\widehat{\bar V}, \notag\\
 h&=\widehat h+iL_-, \quad Y=\widehat Y+F+cL_+
\end{align}
where the hatted fields are periodic. Note that the compact boson has an extra winding term controlled by $F$. The coset partition function then becomes: 
\begin{align}
    Z_c=\sum_{\omega,m_X\in\mathbb Z}\int \mathcal{D} \widehat\phi\mathcal{D}\widehat h \mathcal{D} \widehat V\mathcal{D}\widehat{\bar V}\mathcal{D}\widehat Y \int d^2u e^{-S_{\text{gauged}}\left[\phi, h ,V,\bar V, Y\right]}
\end{align}
\subsection{The Ray-Singer Torsion}
As in \cite{Maldacena:2000kv,Hanany:2002ev}, the path integral over $\widehat{V},\widehat{\bar V}$ gives the Ray-Singer analytic torsion \cite{Ray:1973sb}, where the relevant action is: 
\begin{align}
    S_V=\frac{k}{\pi}\int\dd^2z\,
 \left|\left(\bp+\bp\widehat\phi+\frac u{2{\tau_2}}\right)
 \widehat V\right|^2
\end{align}
Writing $D_u=\bp+u/(2{\tau_2})$, the $\widehat \phi$ field decouples under a chiral rotation $e^{-\widehat\phi}D_ue^{\widehat\phi}$ \cite{Gawedzki:1988nj,Maldacena:2000kv}. Thus, the path integral over these fields yields: 
\begin{align}
     Z_V\propto\exp\!\left[\frac2\pi\int
     d^2z\,\partial\widehat\phi\bp\widehat\phi\right]\mathcal D_\tau(u)
\end{align}
where the $\widehat \phi$ part combined with the kinetic term gives the familiar shift $k\to\kappa\equiv k-2$, and the analytic torsion is defined as: 
\begin{align}
    \mathcal D_\tau(u)=
 e^{2\pi(\im u)^2/{\tau_2}}
 \left|\frac{\eta(\tau)}{\vartheta_1(\tau,u)}\right|^2
 \label{torsion}
\end{align}
where $\vartheta_1(\tau,u)$ is an elliptic theta function. The periodicity of this determinant, as we shall see, allows us to divide the holonomy plane into strips, as was done in \cite{Hanany:2002ev,Israel:2003ry}.  The ghosts associated with gauging have weight $(h_b,h_c)=(1,0)$, and their determinants simply cancel the remaining scalars' oscillator factors, leaving us: 
\begin{align}
\label{kernel}
     Z_c(\tau)={}&\frac{R\sqrt{\kappa K}}{\sqrt {\tau_2}\,|\eta(\tau)|^2}
 \sum_{\omega,m_X\in\ZZ}\int_{\CC}\frac{\dd^2u}{{\tau_2}}\,
 \mathcal D_\tau(u) \notag \\[-1mm]
 &\hspace{8mm}\times
 \exp\!\left[-\frac\pi {\tau_2}
 \left(k|u|^2+|R(\omega\tau-m_X)+icu|^2\right)\right].
\end{align}
As a consistency check, at $c=0$, the above partition function factorizes into the cigar partition function \cite{Hanany:2002ev} and a compact free boson. Moreover, by coupling the $SL(2,\mathbb R)/U(1)$ coset blocks with a suitably chosen lattice of Cartan charges, one can recover the $SL(2,\mathbb R)$ WZW model, with the appropriate charge identifications \cite{Israel:2003ry}. In spirit, this operation is similar to our construction, where by starting with $(SL(2,\mathbb R)\times U(1))/U(1)$, the gauging couples the additional $U(1)$ to the $SL(2,\mathbb R)$, which becomes ungauged in the $c\to \infty$ limit. In order to obtain the one-loop amplitude of a critical bosonic theory, we need to include the internal CFT and the usual $bc$ ghosts: 
\begin{equation}
 \mathcal A_1=\frac12\int_{\mathcal F}\frac{\dd^2\tau}{{\tau_2}^2}
 \,{\tau_2}|\eta(\tau)|^4 Z_{\mathrm{int}}(\tau)Z_c(\tau).
 \label{tor:amplitude}
\end{equation}
where
\begin{align}
    Z_{\text{int}}(\t,\bar \t)=(q\bar q)^{-\frac{c_{\mathrm{int}}}{24}}\sum_i q^{h_i} \bar q^{\bar h_i}, \quad q=e^{2\pi i \t}
\end{align}
We require the internal CFT to have central charge $
c_{\mathrm{int}}=26-\frac{3k}{k-2}$ so that the total central charge of the theory vanishes.

\section{The worldsheet spectrum}
\label{section4}
We now move on to study the worldsheet spectrum of the coset theory. We follow the holonomy techniques of \cite{Hanany:2002ev}, extending them to the charged coset. Readers familiar with these techniques might wish to skim this section and skip to the final result. The finite-deformation partition function was already constructed in \cite{Israel:2003ry}, with explicit conformal weight shifts expanded to first order around the undeformed limit. It is useful for us to perform the decomposition exercise in full, in view of the later studies of the spacetime spectrum for the horizonless geometry and the worldsheet spectrum of the two dimensional charged black hole. We record in this section the explicit character decomposition of this coset at finite deformation, as well as the remainder terms from the full oscillator content \cite{Israel:2004ir, Eguchi:2010cb}.

Since $u\in\mathbb C$, we write $u=x+iy$, and Poisson resum the relevant $m_X$ terms in (\ref{kernel}): 
\begin{align}
\sum_{m\in\mathbb Z}
e^{-\frac{\pi}{\tau_2}|R(\omega\tau-m)+icu|^2}=\frac{\sqrt{\tau_2}}{R}\sum_{n\in\mathbb Z}
\exp\!\left[-\frac{\pi\tau_2 n^2}{R^2}+2\pi in\omega\tau_1-\frac{2\pi icn}{R}y
-\frac{\pi}{\tau_2}(R\omega\tau_2+cx)^2\right].
\end{align}
As discussed above, the periodicity of the analytic torsion allows us to decompose the holonomy in the following strips \cite{Hanany:2002ev,Israel:2003ry}:
\begin{align}
    u=(\nu+\s)\tau-b_0, \quad \nu\in \mathbb{Z}, \quad 0<\s<1, \quad b_0\in \mathbb R
\end{align}
Moreover, the same periodicity $u\to u+\tau \nu$ allows us to define $U=u-\nu \t=\s \t-b_0$, and write $\mathcal D_\tau(u)=\mathcal D_\tau(U)$. The complex parametrization of $u$ now becomes: $y=\tau_2(\nu+\s)$, $x=(\nu+\s)\tau_1-b_0$, and the holonomy measure now becomes:
\begin{align}
    \frac{dxdy}{\t_2}=dxd\s
\end{align}
To perform the holonomy integrals, we need to expand the torsion. Defining $q=e^{2\pi i \tau}$, we can expand the analytic torsion (\ref{torsion}) as \cite{Israel:2004xj,Israel:2004ir}: 
\begin{align}
\label{torsionexpand}
\mathcal D_\tau(U)=\frac{e^{2\pi\tau_2(\sigma^2-\sigma)}}{|\eta(\tau)|^4}
\sum_{a,\bar a\in\mathbb Z}S_a(q)S_{\bar a}(\bar q)  \times e^{-2\pi\tau_2\sigma(a+\bar a)}
e^{2\pi i(x-\nu\tau_1)(a-\bar a)},  
\end{align}
where $S_a(q)=\sum_{b\ge0}(-1)^bq^{b(b+2a+1)/2}$ and $a$ counts the number of $J^+_{r\leq 0}$ minus the number of $J^-_{r<0}$ operators. We also define $a-\bar a=L\in \mathbb Z$. The worldsheet partition function is now given by: 
\begin{align}
Z_c={}&\frac{\sqrt{\kappa K}}{\abs{\eta}^2}
\sum_{\nu,n,\omega\in\mathbb Z}\int_{\mathbb R}dx
\int_0^1d\s\,\mathcal D_\tau(U)\notag\\
&\times\exp\!\left[-\frac{\pi}{\tau_2}(Kx^2+ky^2)
-\pi\tau_2\left(\frac{n^2}{R^2}+R^2\omega^2\right)-2\pi cR\omega x
-\frac{2\pi icn}{R}y+2\pi in\omega\tau_1\right].
\end{align}
It is now easy to perform the holonomy integral over $x$: 
\begin{align}
    I_x=&\int_{\mathbb R}dx\exp\left[-\frac{\pi}{\tau_2}Kx^2-2\pi x(cR\omega-iL)\right] \notag \\
    =&\sqrt{\frac{\t_2}{K}}\exp\left[\frac{\pi \t_2}{K}\p{cR\omega-iL}^2\right]
\end{align}
where we have simply completed the square and performed the Gaussian integral. The second holonomy integral is far less trivial. Collecting the terms that depend on $\s$ yields: 
\begin{align}
    \exp\left[-\pi \t_2\kappa \s^2-2\pi \t_2 \s B\right], \quad B=1+2a-L+k\nu+i\chi
\end{align}
where we have defined $\chi=cn/R$. The second holonomy integral is thus given by: 
\begin{align}
I_\s(B)&=\sqrt{\kappa\tau_2}\int_0^1d\s\,
e^{-\pi\tau_2(\kappa\s^2+2B\s)},\notag\\
Z_c(\tau)&=\sum_{\nu,L,n,\omega\in\mathbb Z}
\sum_{a\in\mathbb Z}
\frac{F_\nu}{|\eta|^6}S_a(q)S_{a-L}(\bar q)\,I_\s(B),
\qquad \bar a=a-L
\end{align}
and $F_\nu$ is the rest of the exponent and the full form is given in (\ref{exponent1}).
\subsection{Discrete representation}
The above integral over $\s$ can be performed in closed form, but to isolate the contribution from discrete and continuous representations, it is useful to introduce an auxiliary parameter and write the integral as \cite{Maldacena:2000kv,Hanany:2002ev}: 
\begin{align}
    I_\s=2\t_2\int_0^1d\s \int_\mathbb R ds e^{-4\pi \t_2 s^2/\kappa} e^{2\pi \t_2 (2is-B)\s}
\end{align}
While introduced as an auxiliary integration variable, $s$ actually has a physical interpretation as the momentum of the Liouville field,
which is coupled to other degrees of freedom of the theory \cite{Israel:2003ry}. The integral over $\s$ is now elementary, and we obtain: 
\begin{align}
    I_\s=\frac{1}{\pi}\int_{\mathbb R}ds e^{-4\pi \t_2 s^2/\kappa}\left[\frac{e^{2\pi \t_2(2is-B)}}{2is-B}-\frac{1}{2is-B}\right]
\end{align}
Focusing on the total exponent of the first term, we can complete the square: 
\begin{align}
    -\frac{4\pi \t_2}{\kappa}\p{s-\frac{i\kappa}{2}}^2-\pi \t_2(\kappa+2B)
\end{align}
Note that the Gaussian is centered on $s=i\kappa/2$, and it is natural to shift the contour up \cite{Maldacena:2000kv,Hanany:2002ev}: 
\begin{align}
\label{sintegral}
    \int_\mathbb R ds f_1(s)=\int_{\mathbb R+i\kappa/2}dsf_1(s)+2\pi i  \sum_{\text{crossed poles}}\text{Res}f_1
\end{align}
The pole is located at $s_*=-iB/2$, and contributes a residue when: 
\begin{align}
\label{contour}
 \frac{1}{2i}e^{\pi \t_2 B^2/\kappa},\qquad   0<\Im s_*<\frac{\kappa}{2}
\end{align}
These poles picked up by the shifted contour give the contribution to the discrete representation, while the continuous representation hides in the remaining integral. If the pole lies exactly on $\Im s=0,\kappa/2$, we will count it with half of the usual contribution, following \cite{Israel:2004xj}. To recover the full vacuum factor, we expand the $\abs{\eta}^{-2}$ term in (\ref{kernel}), and combining with the $\abs{\eta}^{-4}$ from the torsion gives: 
\begin{align}
   \frac{S_a(q)S_{\bar a}(\bar q)}{|\eta|^6}
    =(q\bar q)^{-1/8}\bigl[P(q)^3S_a(q)\bigr]\bigl[P(\bar q)^3S_{\bar a}(\bar q)\bigr],
\end{align}
where $P(q)=\prod_{r=1}^{\infty}(1-q^r)^{-1}$. The $\frac{1}{8}$ is the vacuum power and combining this with the residue should be matched to:
\begin{align}
  h-\frac{c_{\mathrm{ws}}}{24}
   &=-\frac18-\frac{B^2}{4\kappa}
   \implies h=\frac{1-B^2}{4\kappa}
\end{align}
and the worldsheet central charge is $c_{\mathrm{ws}}=3k/\kappa$. With $h=\frac{j(1-j)}{k-2}$, we can solve for $B$ with the choice $\re B<0$: 
\begin{align}
    B=1-2j.
\end{align}
At $c=0$, the above condition is simply the gauge constraint $k\nu+1+a+\bar a=1-2j$ of \cite{Hanany:2002ev}, and the condition (\ref{contour}) becomes the usual discrete window \cite{Maldacena:2000hw}: 
\begin{align}
    \frac{1}{2}<\re j<\frac{k-1}{2}
\end{align}
The eigenvalue of $J_0^3$ is given by $m=j+a$, and similarly $\bar J_0^3$ with $\bar m=j+\bar a$. Using $j=(1-B)/2$ and $a-\bar a=L$, we can solve these as: 
\begin{align}
    m=\frac{L-k\nu-\frac{icn}{R}}{2},\qquad \bar m=\frac{-L-k\nu-\frac{icn}{R}}{2}
\end{align}
Again, in the $c=0$ limit these are simply the familiar charges of the $SL(2,\mathbb R)/U(1)$ coset.  With these, we can now move on to compute the conformal weights of this coset theory. Collecting the rest of the exponent: 
\begin{align}
\label{exponent1}
F_\nu={}&\exp\!\left[-\pi\tau_2\left(
\frac{kR^2\omega^2+L^2+2iLR\omega c}{K}
+k\nu^2+\frac{n^2}{R^2}+2i\chi\nu\right)\right]\notag\\
&\times\exp\!\left[2\pi i\tau_1(n\omega-\nu L)\right]
=q^{H_L}\bar q^{H_R}.
\end{align}
we can solve for $H_{L,R}$ as: 
\begin{align}
\label{Hweights}
    H_L=\frac{m^2}{k}+\frac{P_L^2}{4}, \qquad H_R=\frac{\bar m^2}{k}+\frac{P_R^2}{4}
\end{align}
where
\begin{align}
P_L=\sqrt{\frac Kk}\frac nR+\sqrt{\frac kK}\,R\omega+\frac{icL}{\sqrt{kK}},
\quad P_R=\sqrt{\frac Kk}\frac nR-\sqrt{\frac kK}\,R\omega-\frac{icL}{\sqrt{kK}}
\end{align}
and the conformal weights are: 
\begin{align}
\label{csweight}
    h_{\text{cs}}=-\frac{j(j-1)}{k-2}+H_L, \qquad \bar h_{\text{cs}}=-\frac{j(j-1)}{k-2}+H_R
\end{align}
with $\frac{1}{2}<\re j<\frac{k-1}{2}$. Note that in the neutral limit $c=0$, the above weights simply become the ones obtained in \cite{Hanany:2002ev} coupled to a free boson. Moreover, the weights we have obtained here are not real, due to the terms involving $ic$. This can be traced back to the Euclidean nature of the target space geometry. In the previous computation, we have chosen to gauge the spacelike $U(1)$ subgroup, which yields a Lorentzian black string geometry. After the Wick rotation however, the $B$-field becomes complex. If we had chosen to gauge a compact subgroup, the spacetime geometry is Euclidean, and we would have obtained the real weights. Indeed when we require $c'=ic\in \mathbb R$, the weights all become real, but this would change the integration contour and the pole structure. Thus, we cannot interpret these representations as literally the discrete representation, but some analytic continuation of sorts. This however, will not affect the spacetime spectrum, which remains real after Wick rotation. 

\subsection{Continuous representation}
The continuous series representation hides in the remaining integral of (\ref{sintegral}): 
\begin{align}
\label{contint}
I_\s^{\mathrm{Rem}}(B)
={}&\frac1\pi\int_{\mathbb R}ds\,
e^{-4\pi\tau_2s^2/\kappa}
\left[\frac{e^{-\pi\tau_2(\kappa+2B)}}{2is-B-\kappa}
-\frac1{2is-B}\right]
\end{align}
where we have performed a change of integration variable of the first integral $x=s-i\kappa/2$ and relabeled $x\to s$. It carries the same oscillator factors and prefactors as the original holonomy integral: 
\begin{align}
    Z_{c}^\mathrm{cont}(\tau)=\sum_{\nu,L,n,\omega\in\mathbb Z}
\sum_{a\in\mathbb Z} \frac{F_\nu}{|\eta|^6}S_a(q)S_{a-L}(\bar q) I_\s^{\mathrm{Rem}}(B)
\end{align}
As in the discrete representation, we include the $\p{q\bar q}^{-1/8}$ coming from the $|\eta(\tau)|^{-6}$ expansion and combine it with the common exponent: 
\begin{align}
    -\frac{1}{8}+\frac{s^2}{\kappa}=-\frac{c_{\text{ws}}}{24}+\frac{s^2+\frac{1}{4}}{\kappa}
\end{align}
Indeed, this gives the expected weight for the continuous representation where $j=\frac{1}{2}+is$. Including all the exponent as before, we can read off the conformal weight to be: 
\begin{align}
    h^{\text{cont}}_{\text{cs}}=\frac{s^2+\frac{1}{4}}{k-2}+H_L, \qquad \bar h^{\text{cont}}_{\text{cs}}=\frac{s^2+\frac{1}{4}}{k-2}+H_R
\end{align}
where $H_{L,R}$ are given in (\ref{Hweights}). Again, at $c=0$, the continuous weights reduce to those of \cite{Hanany:2002ev} with a free compact boson. To compute the density of the continuous series states, it is useful to use the following identity: 
\begin{align}
    F_{\nu}e^{-\pi \t_2(\kappa+2B)}=F_{\nu+1}q^a\bar q^{\bar a}
\end{align}
where we defined $F_{\nu}$ in (\ref{exponent1}). We can view this operation as spectrally flowing the states as $\nu \to \nu+1$. Thus, including the exponent and the oscillator expansions, the first term in the integral can be rewritten using the above identity as: 
\begin{align}
\frac{F_\nu}{|\eta|^6}S_a(q)S_{\bar a}(\bar q)
\frac{e^{-\pi\tau_2(\kappa+2B)}}{2is-B-\kappa}=\frac{F_{\nu+1}}{|\eta|^6}
\frac{\bigl[q^aS_a(q)\bigr]\bigl[\bar q^{\bar a}S_{\bar a}(\bar q)\bigr]}{2is-B-\kappa}
\end{align}
Introducing new labels as $\nu'=\nu+1$, $a'=a-1$, and $L'=L$ and similarly for $\bar a'=\bar a-1$, we can rewrite the pole parameter as $B'=B+\kappa$. This indeed can be interpreted as the spectral flow operation performed in \cite{Hanany:2002ev}, which relies on the isomorphism $\widehat{\mathcal D}^{+,w-1}_{j}\simeq \widehat{\mathcal D}^{-,w}_{k/2-j}$. In (\ref{contint}), we can then combine the first term from the $\nu'=\nu+1$ sector and the second term from the $\nu$ sector as: 
\begin{align}
&\sum_{\nu,L}\sum_a \frac{F_\nu}{|\eta|^6}
\frac{q^{a+1}S_{a+1}(q)\bar q^{a-L+1}S_{a-L+1}(\bar q)
-S_a(q)S_{a-L}(\bar q)}{2is-B}\notag\\
&\qquad=\sum_{\nu,L}\sum_a \frac{F_\nu}{|\eta|^6}
\frac{1-S_a(q)-S_{a-L}(\bar q)}{2is-B}
\end{align}
where we have used the identity $q^{r+1}S_{r+1}(q)=1-S_r(q)$. To compute the continuous density, let's introduce: 
\begin{align}
    f_x(s)=\frac{x}{\pi(x^2+s^2)}, \qquad \frac B2=\alpha+a
\end{align}
Thus, we can restrict our integration range to $s\geq 0$ and write: 
\begin{align}
    &\frac1\pi\left[\frac1{2is-B}+\frac1{-2is-B}\right] =-\frac{2B}{\pi(B^2+4s^2)}=-f_{\alpha+a}(s)
\end{align}
Including the multiplicities, the worldsheet continuous density can then be written as: 
\begin{align}
    \rho_{\mathrm{ws}}(s;q,\bar q)=\sum_a \bigl[S_a(q)+S_{a-L}(\bar q)-1\bigr]f_{\alpha+a}(s)
\end{align}
As explained in \cite{Maldacena:2000kv,Hanany:2002ev}, when performing the sum over states $(J_0^+\bar J_0^+)^r\ket{\psi}$, one encounters a logarithmically divergent sum. Indeed, performing the regulated sum yields the familiar density of states with $N=\bar N=0$, which keeps terms like $[q^0]S_a(q)=\mathbf1_{a\ge0}$\footnote{The notation here simply means the coefficient of $q^0$ after expanding $S_a(q)$.}: 
\begin{equation}
\begin{aligned}
\rho_{0,J}(s)
&=\sum_{a=-J}^{J}
\bigl[\mathbf1_{a\ge0}+\mathbf1_{a\ge L}-1\bigr]f_{\alpha+a}(s)\\
&=\frac{2\log J}{\pi}
+\frac1{2\pi i}\partial_s\log
\frac{\Gamma(\alpha-is)\Gamma(1-\bar\alpha-is)}
{\Gamma(\alpha+is)\Gamma(1-\bar\alpha+is)}+o(1),
\qquad J\longrightarrow\infty
\end{aligned}
\label{gammadensity}
\end{equation}
where we have introduced $J$ as the cutoff parameter to regulate the divergent sum, and in the end we take the $J\to \infty$ limit. We also have:
\begin{align}
\alpha=\frac12-m,\qquad
\bar\alpha=\frac12-\bar m=\alpha+L.
\end{align}
The divergent $\log J$ is an IR volume divergence, which comes from the long strings being able to reach the asymptotic boundary. Retaining the full oscillator multiplicity yields an additional remainder density $\mathcal R(s;q,\bar q)$ that does not belong to the usual highest weight representations of $SL(2,\mathbb R)$. Similar contributions are discussed in \cite{Israel:2004ir,Eguchi:2010cb}, whose physical interpretation is not straightforward. However, they are not made explicit in the analysis of \cite{Hanany:2002ev}. In our notation, this term is given by: 
\begin{align}
\label{finitereminder}
    \mathcal R(s;q,\bar q)=\sum_{a\in\mathbb Z} \bigl[S_a(q)-\mathbf1_{a\ge0}\bigr]f_{\alpha+a}(s)+\sum_{a\in\mathbb Z}
\bigl[S_a(\bar q)-\mathbf1_{a\ge0}\bigr]f_{\bar\alpha+a}(s)
\end{align}
The full regulated continuous density can now be written as:
\begin{align}
\rho_{\mathrm{ws},J}(s;q,\bar q)
={}&\sum_{a=-J}^{J}
\bigl[S_a(q)+S_{a-L}(\bar q)-1\bigr]f_{\alpha+a}(s)\notag\\
={}&\rho_{0,J}(s)+\mathcal R(s;q,\bar q)+o(1),
\qquad J\longrightarrow\infty,
\end{align}
Inserting this into the remaining $s$ integral then gives the continuous series contribution. We have thus recovered the expected character decomposition from the worldsheet partition function. With these results, we can move on to investigate the spacetime spectrum. 

\section{Spacetime spectrum}
\label{section5}
In this section, we study the spacetime spectrum of the coset model. Our motivation is quite simple: we would like to recover the MOS spectrum in the ungauged limit of the coset model by directly evaluating the holonomy and modular integrals. As discussed in section \ref{Section2}, the natural coordinate to unwind as the Lorentzian time for this purpose is the compact $X$ direction, which yields a horizonless geometry associated with the $J_3\bar J_3$ deformed $SL(2,\mathbb R)$ theory \cite{Israel:2003ry,Chakraborty:2024mls}, which in the ungauged limit simply becomes thermal AdS$_3$. On the other hand, if we choose to unwind the $\theta$-direction, the corresponding geometry is that of the Horne--Horowitz black string. However, as we shall see, it is not straightforward to extract its spacetime spectrum from the Euclidean partition function, and we will mainly focus on the horizonless geometry. First we recover the spacetime trace of a single string excitation from the worldsheet results. Next, we directly evaluate the holonomy and unfolded modular integral, and demonstrate that it reconstructs the spacetime excitation trace. Finally, we comment on the Horne--Horowitz continuation and its spacetime spectrum. Throughout this section we take $c>0$, $k>3$, and $(h_I,\bar h_I)$ are the weights of the internal CFT.

\subsection{Spacetime spectrum from spectral decomposition}
\label{section5.1}

Let's start from the worldsheet spectrum and continue the $X$-direction as our Lorentzian time. In this continuation, both the holonomy and modular integrals can be performed, and one can extrapolate the spacetime spectrum given by that of the $J_3\bar J_3$-deformed AdS$_3$ string theory \cite{Chakraborty:2024mls}. Concretely, we write $x=\frac{X}{c}=it_x$ as our Lorentzian time direction. In this case we would like to obtain the Lorentzian mode $e^{-iEt_x}$, which amounts to continuing: 
\begin{align}
    \frac{cn}{R}\to iE
\end{align}
as our spacetime energy. Plugging the above continuation into (\ref{csweight}) and including the oscillator levels, we obtain the Lorentzian conformal weights, which are now all real: 
\begin{align}
h_{\rm cs} &=-\frac{j(j-1)}{\kappa} -\nu m-\frac{k\nu^2}{4}
+N+\frac{L^2}{4K}-\frac{E^2}{4c^2},\\
\bar h_{\rm cs} &=-\frac{j(j-1)}{\kappa} -\nu\bar m-\frac{k\nu^2}{4} +\bar N+\frac{L^2}{4K}-\frac{E^2}{4c^2}.
\end{align}
where we have set $\omega=0$ (see below (\ref{eqn5.17})). The charges are also real:
\begin{align}
   m=\frac{E-k\nu+L}{2},\qquad\bar m=\frac{E-k\nu-L}{2}.
\end{align}
After imposing the mass-shell condition and choosing the positive energy branch, we obtain the spacetime energy for the continuous series representation with $j=\frac{1}{2}+is$: 
\begin{align}
\label{cosetcontenergy}
E_{c,\nu}(s)=c\sqrt{
K\nu^2+2\mathcal N-A_k+\frac{L^2}{K}+\frac{4s^2}{\kappa}}-c^2\nu, \quad 
A_k=4-\frac1{\kappa}
\end{align}
where $\mathcal N=N+\bar N+h_I+\bar h_I$. In the ungauged limit $c\to \infty$, and for positively flowed states $\nu>0$, the above energy becomes the familiar continuous series energy relation in AdS$_3$ \cite{Maldacena:2000hw}. For the discrete series representation, the energy $E$ and the spin $j$ are not independent. Using the discrete pole condition $B=1-2j$, they are tied by the constraint: 
\begin{align}
E=2j+k\nu+a+\bar a, \qquad L=a-\bar a
\end{align}
After imposing the mass-shell condition, we can then eliminate $j$ and solve for the energy, where the positive branch is: 
\begin{align}
\label{discrete-energy}
E_{\mathrm{dis}}=\frac{c^2}{c^2+\kappa}\left[d-\kappa\nu+\sqrt{(d-\kappa\nu)^2-\left(1+\frac{\kappa}{c^2}\right)
\left[d^2-\kappa\left(
k\nu^2+2\mathcal N-4+\frac1\kappa+\frac{(a-\bar a)^2}{K}\right)\right]}\right]
\end{align}
and $d=1+k\nu +a +\bar a$. The discrete spin window now becomes: 
\begin{align}
         d<E_{\mathrm{dis}}<d+\kappa
\end{align}
As a consistency check, taking the $c\to \infty$ limit recovers the discrete energy in thermal AdS$_3$: 
\begin{align}
  E_{\mathrm{dis}}^{(\infty)}=1+a+\bar a+2\nu+\sqrt{1+4\kappa\left(N_\nu+h_I-1-\frac{\nu(\nu+1)}2\right)}
\end{align}
together with the spin: 
\begin{align}
\frac{1}{2}<j_\infty=\frac12\left[1-\kappa\nu+\sqrt{1+2\kappa\left[\mathcal N-2-\nu(\nu+1+a+\bar a)\right]}\right]<\frac{k-1}{2}
\end{align}
where we have introduced back the flowed oscillator levels $N_\nu=N-\nu a,\bar N_\nu=\bar N-\nu\bar a$, and used the level matching condition.

To obtain the continuous density from the worldsheet, we need to include the ghost oscillators $|\eta|^4$, which changes the $|\eta|^{-6}$ to $|\eta|^{-2}$. Dropping the vacuum power, which was already grouped to give the vacuum energy, we have the following expansion: 
\begin{align}
P(q)=\prod_{r=1}^{\infty}(1-q^r)^{-1}
=\sum_{N=0}^{\infty}p_Nq^N,
\qquad
P(q)S_a(q)=\sum_{N=0}^{\infty}d_+(N,a)q^N
\end{align}
Thus, at fixed $(N,\bar N)$, the coefficient of the expansion for $q^N \bar q^{\bar N}$ is: 
\begin{align}
[q^N\bar q^{\bar N}]
\bigl[P(q)P(\bar q)\rho_{\mathrm{ws},J}\bigr]
=-\sum_{a=-J}^{J} \mathcal D_{\mathrm {total}}
f_{\alpha+a}(s),\\
\mathcal D_{\mathrm {total}}
=p_Np_{\bar N}
-p_{\bar N}d_+(N,a)
-p_Nd_+(\bar N,a-L)
\end{align}
where the minus sign follows from the definition of $\mathcal D_{\mathrm{total}}$, and this sign will be canceled below due to $f_x(s)$ being odd in $x$. Moreover, the mass-shell condition makes the continuous energy dependent on $s$, i.e. $E=E(s)$, and we have: 
\begin{align}
x_a(s)&=\frac{E_{c,\nu}(s)-k\nu+L-1}{2}-a, \notag \\
f^{\mathrm{on}}_{x_a}(s)
&=\frac{1}{2\pi i}\frac{d}{ds}
\log\frac{x_a(s)+is}{x_a(s)-is}=\frac{x_a(s)-\frac{s}{2}E'_{c,\nu}(s)}
{\pi\bigl[x_a(s)^2+s^2\bigr]}
\end{align}
This then gives the continuous density of states for the spacetime spectrum: 
\begin{align}
\label{condensws}
\rho_{N,\bar N}^{J}(s|\nu,L,I)
=\sum_{a=-J}^{J}\mathcal D_{\mathrm{total}} f_{x_a}^{\mathrm{on}}(s)
\end{align}
which we will reproduce again directly from the modular integral in section \ref{section5.2}. Now that we have the spacetime energy and density, we can assemble the single-string excitation trace using the continuous density we obtained from the worldsheet (\ref{condensws}): 
\begin{align}
z_c(b)
={}&
\sum_{\nu,N,\bar N,a,\bar a,I}
d_I\,d_+(N,a)d_+(\bar N,\bar a)
\,e^{-bE_{\mathrm{dis}}} +
\sum_{\nu,L,N,\bar N,I}
d_I
\int_0^\infty ds\,
\rho_{N,\bar N}^{J}
e^{-bE_{c,\nu}(s)}
\end{align}
where $b=\frac{2\pi R}{c}$ is the thermal period, which is the same parameter we used as the inverse temperature $b=2\pi\ell$ for thermal AdS$_3$. The discrete sum is restricted to the spin window and all states satisfy the level matching condition $\Delta=N+h_I-\bar N-\bar h_I=\nu L$. Here $d_+$ represents the discrete character coefficients at a given level while $d_I$ is for the internal CFT. 

\subsection{Evaluation of integrals }
\label{section5.2}
Having assembled the single-string excitation trace from the worldsheet spectrum, we would like to recover this directly from the Euclidean path integral. For this we need to include the internal CFT and the ghost fields, and use the $X$-winding sum to unfold the fundamental domain into a strip \cite{Polchinski:1985zf,Maldacena:2000kv}. The holonomy and modular integrals give both the energies for the discrete and continuous representation, as well as the regulated continuous density. As in \cite{Maldacena:2000kv}, the zero winding is kept separate and not included in the amplitude. Here we will only present the main steps and the details of the integrals can be found in Appendix \ref{intappendix}.

Our starting point is the amplitude (\ref{tor:amplitude}), which we reproduce in full here for convenience: 
\begin{align}
\label{eqn5.17}
\mathcal A_{c}
={}&
\frac{R\sqrt{\kappa K}}2 \sum_{\omega ,m_X\in\mathbb Z}\int_{\mathcal F} \frac{d^2\tau}{\tau_2^{5/2}}\, |\eta|^2 Z_{\mathrm{int}} \notag
\\&\times \int_{\mathbb C}d^2u\,\mathcal D_\tau(u)
\exp\!\left[-\frac{\pi}{\tau_2} \left(k|u|^2+|R(\omega \tau-m_X)+icu|^2\right)\right].
\end{align}
First we unfold the fundamental domain by choosing $(\omega,m_X)=(0,-r)$ with $r \in \mathbb Z_{>0}$. The contribution from $r<0$ is identical to that of $r>0$ under $u\to-u$, and combining them cancels the factor $1/2$. Thus, the single cover integral that we need to evaluate is given by: 
\begin{align}
z_{c}(b)
={}&R\sqrt{\kappa K}
\int_0^\infty\frac{d\tau_2}{\tau_2^{5/2}}
\int_{-1/2}^{1/2}d\tau_1\,
|\eta|^2 Z_{\mathrm{int}} \notag
\\ &\times \int_{\mathbb C}d^2u\,\mathcal D_\tau(u)
\exp\!\left[ -\frac{\pi}{\tau_2} \left(k|u|^2+|R+icu|^2\right)\right],
\qquad R=\frac{bc}{2\pi},
\end{align}
To perform the integrals, we again divide the holonomy into strips: 
\begin{align}
\label{strips5}
u=(\nu+\sigma)\tau-b_0, \qquad \nu\in\mathbb Z,\quad 0<\sigma<1,\quad b_0\in\mathbb R.
\end{align}
The measure is $d^2u=\tau_2\,d\sigma\,db_0$, and the periodicity of the torsion $U=\s \t-b_0$ gives $\mathcal{D}_\t(U)$. Expanding the oscillators, the analytic torsion, and the classical Gaussian yields (see Appendix \ref{intappendix}): 
\begin{align}
\label{totalexpand}
    z_c(b)
={}&R\sqrt{\kappa K}\sum_{\nu,N,\bar N,a,\bar a,I}d_I\,d_+(N,a)d_+(\bar N,\bar a) \notag
\\&\times\int_0^\infty\frac{d\tau_2}{\tau_2^{3/2}}\int_{-1/2}^{1/2}d\tau_1\int_0^1d\sigma\int_{\mathbb R}db_0
\notag \\&\times\exp\!\left[-\frac{\pi R^2}{\tau_2}+2\pi cR(\nu+\sigma)-\frac{\pi K}{\tau_2}\bigl[b_0-(\nu+\sigma)\tau_1\bigr]^2\right]
\notag \\ &\times\exp\!\left[-\pi\tau_2\left(K(\nu+\sigma)^2-2\sigma^2+2\sigma(1+a+\bar a)+2\mathcal N-A_k\right)\right]
\notag \\&\times\exp\!\left[2\pi i\tau_1(\Delta+\sigma L)-2\pi ib_0L\right]
\end{align}
where $A_k$ is defined in (\ref{cosetcontenergy}), and we also defined
\begin{align}
\mathcal N=N+\bar N+h_I+\bar h_I,
\qquad \Delta=N+h_I-\bar N-\bar h_I,\qquad L=a-\bar a
\end{align}
Integrating over $b_0$ yields $\sqrt{\tau_2/K}\,e^{-\pi\tau_2L^2/K}e^{-2\pi i(\nu+\sigma)\tau_1L}$, and combining this factor with the $\t_1$ phase already present in (\ref{totalexpand}) gives: 
\begin{align}
    \int_{-1/2}^{1/2}d\tau_1\,e^{2\pi i\tau_1(\Delta-\nu L)}=\delta_{\Delta,\nu L}.
\end{align}
and integrating over $\tau_1$ is simply the level matching condition. Collecting the relevant terms, the integral over $\t_2$ is:
\begin{align}
    I&(\mathcal Q) =\int_0^\infty\frac{d\tau_2}{\tau_2}
    \exp\!\left[-\frac{\pi R^2}{\tau_2}-\pi\mathcal Q(\s)\tau_2\right], \notag \\
    &\mathcal Q_{c,\nu}(\sigma) =K(\nu+\sigma)^2-2\sigma^2 +2\sigma(1+a+\bar a)+2\mathcal N-A_k+\frac{L^2}{K}
    \label{definitionofA}
\end{align}
Note that if $\mathcal Q(\s)>0$, the above integral can be performed exactly and is simply a modified Bessel function. On the other hand, for $\mathcal Q(\s)<0$, the above integral diverges exponentially as $\tau_2\to \infty$, and produces a tachyonic infrared divergence \cite{Maldacena:2000kv}. We will assume $\mathcal Q(\s)>0$ and treat the other branch by analytical continuation. Ignoring for now the multiplicities, the cell contribution is given by: 
\begin{align}
    z_{\mathrm{cell}}(b)=\frac{bc\sqrt{\kappa}}{\pi}\int_0^1d\sigma\,
e^{bc^2(\nu+\sigma)}
K_0\!\left(bc\sqrt{\mathcal Q_{c,\nu}(\sigma)}\right)
\label{zcellbessel}
\end{align}
To extract the spectrum, we use the integral representation of the Bessel function: 
\begin{align}
    z_{\mathrm{cell}}(b)=\frac{bc\sqrt{\kappa}}{\pi}\int_0^1d\sigma\int_{c\sqrt{\mathcal Q(\sigma)}}^\infty d\varepsilon \hspace{1mm} \frac{e^{-b[\varepsilon-c^2(\nu+\sigma)]}}{\sqrt{\varepsilon^2-c^2\mathcal Q_{c,\nu}(\sigma)}}.
\end{align}
Note that the exponent becomes $-bE$ if we define $E=\varepsilon-c^2(\nu+\sigma)$, and the cell contribution can be written as: 
\begin{align}
z_{\mathrm{cell}}(b)&=b\int_{E_0}^{\infty}dE\,e^{-bE}\,
\mathscr N_{\mathrm{cell}}(E), \notag \\
\mathscr N_{\mathrm{cell}}(E)&=\frac{\sqrt{\kappa}}{\pi}
\int_{\substack{0\leq\sigma\leq1\\E+c^2(\nu+\sigma)\geq
c\sqrt{\mathcal Q_{c,\nu}(\sigma)}}}\frac{d\sigma}{\sqrt{\frac{[E+c^2(\nu+\sigma)]^2}{c^2}-\mathcal Q_{c,\nu}(\sigma)}}
\label{zcellE}
\end{align}
Note that positivity inside the square root then selects the positive branch: 
\begin{align}
    E+c^2(\nu+\s)\geq c\sqrt{\mathcal Q_{c,\nu}(\s)}
\end{align}
while the range of $\s$ integration is from $0\leq \s\leq 1$. The lower limit of the $dE$ integral is thus given by: 
\begin{align}
\label{minenergy}
      E_0=E_{\min}^{\mathrm{cell}} =\min_{0\leq\sigma\leq1}\left[c\sqrt{\mathcal Q_{c,\nu}(\sigma)}-c^2(\nu+\sigma)\right]
\end{align}
The integral in (\ref{zcellE}) already exhibits the form of a thermal trace, but to make it manifest we can integrate it by parts:
\begin{align}
\label{intbypartmain}
b\int_{E_0}^{\infty}dE\,e^{-bE}\mathscr N_{\mathrm{cell}}(E)
={}&e^{-bE_0}\mathscr N_{\mathrm{cell}}(E_0)+\int_{(E_0,\infty)}
e^{-bE}\,d\mathscr N_{\mathrm{cell}}(E).
\end{align}
This manifests the structure of the discrete and continuous states, and our goal now is to compute the respective energy and verify this is indeed the case. 
\subsubsection{Discrete states}
Here we focus on the first term that will turn out to be the discrete series contribution. Assume the minimum energy (\ref{minenergy}) of the discrete series occurs within the range $0<\s_*<1$. Minimizing the energy then gives (see Appendix \ref{intappendix}): 
\begin{align}
\label{sigmastar}
    \sigma_*=\frac{E_0-d}{\kappa}
\end{align}
If we define the spin $j$ of the state through the discrete charge relation: 
\begin{align}
    E_0=2j+k\nu+a+\bar a
\end{align}
we then have $E_0=2j-1+d$, and the requirement that the minimum lies in the $\s$-cell then becomes the correct bound for the discrete spin: 
\begin{align}
0<\sigma_*<1 \quad \implies \quad \frac12<j<\frac{k-1}{2}.
\end{align}
We also need to compute the energy equation. Using equations (\ref{minenergy}), (\ref{sigmastar}), and $E_0-d=2j-1$ then gives:
\begin{align}
   \frac{E_0^2}{c^2}+2\nu E_0=k\nu^2+2\mathcal N-4-\frac{4j(j-1)}{\kappa}+\frac{L^2}{K}.
\end{align}
which is precisely the discrete state energy we obtained from continuing the worldsheet answer. Solving for $E_0$ and keeping the positive branch yields (\ref{discrete-energy}). The integral over $\s$ can be performed within the range $\s_-\leq\s\leq\s_+$, which simply gives $\mathscr N_{\mathrm{cell}}=1$. Recovering the expansion coefficients and the sum then gives precisely the discrete series contribution: 
\begin{align}
  z_{c}^{\mathrm {disc}}=  \sum_{\nu,N,\bar N,a,\bar a,I} d_I\,d_+(N,a)d_+(\bar N,\bar a)\, \delta_{\Delta,\nu(a-\bar a)}\,e^{-bE_{\mathrm{dis}}}
\end{align}

\subsubsection{Continuous states}
Next we turn to the continuous series contribution, which comes from the second integral of (\ref{intbypartmain}). Different from the discrete representations, the two roots $\s_\pm$ can now reach the boundary of the cell, where $\s_-\leq0$ or $\s_+\geq1$, and the energy will start to vary continuously. Denote the endpoints by $e=0,1$ and write the threshold energy as: 
\begin{align}
     E_{\mathrm{th},e}=c\sqrt{\mathcal Q_{c,\nu}(e)}-c^2(\nu+e)
\end{align}
After the energy has surpassed the threshold, we can introduce a continuous parameter $s\geq0$ and write the energy as a function of $s$: 
\begin{align}
    \frac{[E+c^2(\nu+e)]^2}{c^2}
-\mathcal Q_{c,\nu}(e)= \frac{4s^2}{\kappa}
\end{align}
where with a bit of hindsight from the continuous energy we fix the normalization as $4/\kappa$. At the lower point $e=0$, this energy gives the continuous state energy we obtained before: 
\begin{align} 
    E_{c,\nu}(s)=c\sqrt{ K\nu^2+2\mathcal N-A_k +\frac{L^2}{K}+\frac{4s^2}{\kappa}}-c^2\nu
\end{align}
while at $e=1$, the energy is simply the neighboring spectrally flowed sector with
\begin{align}
    \nu'=\nu+1,\quad N'=N+a,\quad \bar N'=\bar N+\bar a, \quad a'=a-1, \quad \bar a'=\bar a-1 
\end{align}
The two energies yield two densities which can be shown to be equivalent under spectral flow, while the multiplicities do in fact change under spectral flow. One can then massage the integral over $d\mathscr N_{\mathrm{cell}}$ into the expected continuous density and impose the common cutoff $-J<a<J$ with the resulting multiplicities. The detailed computation is given in appendix \ref{intappendix}, and the final result of the density is given by: 
\begin{align}
   & \rho_{N,\bar N}^{(J)}(s) =\sum_{a=-J}^{J}\mathcal D_{\mathrm{ total}
    }\,\frac{2\left[E(s)-d_a-s\frac{dE}{ds}\right]}{\pi\left([E(s)-d_a]^2+4s^2\right)}, \qquad d_a=1+k\nu+2a-L, \notag\\
    \mathcal D_{\mathrm {total}}&=d_+(N-a-1,a+1)\, d_+(\bar N-a+L-1,a-L+1)-d_+(N,a)d_+(\bar N,a-L)
\end{align}
The multiplicity matches exactly with (\ref{condensws}), and can be seen by using $q^{a+1}S_{a+1}=1-S_a$, which gives $d_+(N-a-1,a+1)=p_N-d_+(N,a)$. Recovering all the prefactors and expansion coefficients then yields: 
\begin{align}
  z_c^{\mathrm{con} } =\sum_{\nu,L,N,\bar N,I}
d_I\,\delta_{\Delta,\nu L}\int_0^\infty ds\,\rho_{N,\bar N}^{J}e^{-bE_{c,\nu}(s)}
\end{align}
By performing the holonomy and modular integrals, we have successfully recovered the spacetime spectrum of the horizonless geometry \cite{Chakraborty:2024mls}. 

\subsection{The ungauged limit}
\label{section5.3}
As we have shown in the previous section, the ungauged limit $c\to \infty$ correctly produces the energies of the discrete and continuous representations of \cite{Maldacena:2000kv}. There is however a subtle difference in the continuous density limit. Taking the large $c$ expansion of our density yields: 
\begin{align}
    \rho_{N,\bar N,\infty}=p_Np_{\bar N}\rho_{\mathrm{primary},\infty}+\Delta\rho_{N,\bar N,\infty}
\end{align}
where 
\begin{align}
    \rho_{\mathrm{primary},\infty}(s)
={}&\frac{2r_{\mathcal P}}{\pi}+\frac1{2\pi i}\frac d{ds}\log\frac{\Gamma(\frac12-is+\bar m_\infty)\Gamma(\frac12-is-m_\infty)}{\Gamma(\frac12+is+\bar m_\infty)
\Gamma(\frac12+is-m_\infty)}
\end{align}
The first term denotes the prescription dependent $\log$ volume divergence with $r_{\mathcal P}=\log J$ for our prescription. This indeed matches the density found in \cite{Maldacena:2000kv}, but we also have an extra piece of the continuous density $\D\rho_{N,\bar N}$. This term can be traced back to the finite remainder term $\mathcal R_{N,\bar N}$ in (\ref{finitereminder}) \cite{Israel:2003ry,Israel:2004ir}. This is a finite energy dependent shift, rather than a simple shift of the radial volume, and is in general nonzero for $N,\bar N>0$. Tentatively, we believe that these differences might come from the difference in the regularization procedures. Here we used a finite cutoff parameter $J$ to regularize the IR volume divergence, while MOS regulated the long string divergence by a Liouville wall prescription. As discussed explicitly in section 4.3 of \cite{Maldacena:2000kv}, the divergent part depends only on the bulk physics, while the finite remainder part encodes not only the scattering phase shift but also some dependence on this cutoff procedure.

\subsection{Comments on Horne--Horowitz Continuation}
\label{section5.4}
In the case of the HH-black string, we need to continue the $\th$ coordinate as our Lorentzian time, where the asymptotically normalized Euclidean time is given by $t_E=\sqrt{K}\th$. Thus, the black string can be obtained by continuing: 
\begin{align}
    \theta=\frac{it_{HH}}{\sqrt{K}}
\end{align}
while $X\sim X+2\pi R$ remains spatial. For the target space energy, we also need to continue $L \to -i \sqrt{K} E_{HH}$ to obtain a Lorentzian mode $e^{-iE_{HH}t_{HH}}$. Restricting to the unflowed sector $\nu=0$ yields: 
\begin{align}
h_{\mathrm{HH}}
&=-\frac{j(j-1)}{\kappa}+N
+\frac{(P+W_X)^2-(E_{\mathrm{HH}}-\zeta W_X)^2}{4},\\
\bar h_{\mathrm{HH}}
&=-\frac{j(j-1)}{\kappa}+\bar N
+\frac{(P-W_X)^2-(E_{\mathrm{HH}}-\zeta W_X)^2}{4},
\end{align}
where $\zeta=\frac{c}{\sqrt K}$, and we have also defined the spatial momentum and winding as: 
\begin{align}
    P=\frac{n}{R}, \quad W_X=R\omega
\end{align}
The flowed sectors with $\nu\neq0$ do not have a simple Lorentzian continuation due to terms like $i\nu \sqrt{K}E_{HH}$ in the level matching condition, and need to be treated differently. Imposing the mass-shell condition and solving for the positive branch of energy, we obtain: 
\begin{align}
E_{\mathrm{HH}}=\zeta W_X+\sqrt{P^2+W_X^2+2\mathcal N-4-\frac{4j(j-1)}{\kappa}}, \quad E_{HH}-\zeta W_X>0
\end{align}
where $\mathcal N=N+\bar N+h_I+\bar h_I$ as before. For continuous series, we simply take $j=\frac{1}{2}+is$ where $s\in \mathbb R$. The first term in this energy formula can be unpacked as $\zeta R\omega$, which represents the interaction with the background $B$ field. $P$ gives the momentum along the spatial circle while $W_X$ represents the energy of the string stretching in the $X$-direction. Obtaining the complete black string spacetime spectrum is certainly interesting, but other than the aforementioned difficulty in continuing the flowed sectors to Lorentzian geometry, we also need to identify the appropriate thermal interpretation of the modular integral\footnote{See \cite{Hartman:2015lfa,Gillioz:2019iye,Kravchuk:2021kwe,Jafferis:2021ywg,Ashok:2021ffx,Toni:2025nox} for related discussions regarding analytically continuing Euclidean CFT observable to Lorentzian signatures.}. In \cite{Maldacena:2000kv}, the nonzero thermal winding pairs transform as a doublet under the worldsheet modular group $SL(2,\mathbb Z)$, which allows them to unfold the fundamental domain into a strip, and set one of the winding numbers to zero. One then sums over the other winding number over $\mathbb Z_{>0}$. The zero winding $(0,0) $ is treated as the zero temperature vacuum energy and omitted in the amplitude. The gauging in our case does not prevent the unfolding, and as we have seen, the horizonless geometry admits such a procedure. However, for the Horne-Horowitz black string, the $X$ circle remains spatial while we need to unwind the $\theta$ direction, and thus unfolding the same winding sum does not yield a thermal trace with respect to the HH time \cite{Dijkgraaf:1991ba}. Thus, the angular sector must be organized with the holonomies, and the zero holonomy cell cannot be simply discarded as the vacuum energy contribution without establishing its physical interpretation. These additional subtleties need to be clarified when extrapolating the spacetime spectrum for the black string from the Euclidean amplitude. The easier route is perhaps to study the spacetime spectrum directly from the Lorentzian current algebra following \cite{Maldacena:2000hw,Toni:2026vly}.

\section{Near extremality}
\label{section6}
In this section, we turn to another aspect of our studies, which is the near extremal behavior of the target space black strings and black holes. Given that we have an exact worldsheet description of these objects, it is natural to study their quantum effects directly from the worldsheet. This indeed was the route taken by \cite{Toni:2026brc}, where the author investigated the quantum corrections to the two-dimensional charged black hole entropy near extremality. Here, we revisit the problem by identifying two towers of neutral states that are responsible for the $\log T^{3/2}$ correction. The discrete poles appear only below the threshold $T_*=1/(4\pi\sqrt{k})$, and can be taken all the way to extremality. We provide evidence for interpreting them as the worldsheet counterparts of the Schwarzian modes of the AdS$_2$ throat, and discuss the subtleties that must be resolved before establishing this logarithm as a universal correction to the full black hole entropy. For the black string, we use the same procedure to identify two towers of spatially wound string states that yield a $\log T^{2}$ contribution to its one-loop amplitude. 

Here, we will reconstruct the partition function following the holonomy conventions of \cite{Hanany:2002ev}. This procedure closely follows the three-dimensional construction in the previous section, and we will present the main result and leave the details of the construction to the appendix. We will assume $k>3$ throughout this section.

\subsection{Two-dimensional charged black hole}
\label{section6.1}
The construction of the two-dimensional black hole partition function follows similarly to that of the black string, except that the compact boson is now chirally charged, with $c=0$ and $\bar c\in \mathbb R$. Gauge anomaly cancellation requires the constraint  \cite{Quella:2002fk,Giveon:2003ge,Giveon:2004zz}: 
\begin{equation}
k=k\lambda^{\,2}+\bar c^{\,2}.
\end{equation}
and we can solve for $\lambda=\sqrt{1-\bar c^{\,2}/k}$, where $\lambda=1$ yields the uncharged limit. It is also useful to introduce the parameters: 
\begin{equation}
  \xi=\frac{\bar c}{\sqrt{k}(1+\lambda)},
  \qquad \mu =1-\xi^2=\frac{2\lambda}{1+\lambda}
\label{cbhparameters}
\end{equation}
Note that the extremal limit of the black hole is now given by $\abs{\xi}\to 1$ or equivalently $\lambda \to 0$. As before, we can Hodge decompose the gauge field as in (\ref{tor:hodge}), but now the gauging is asymmetric, and we write: 
\begin{equation}
A_z=\partial f+i\partial\widehat h
       -\frac{u_L}{2\tau_2}, \qquad
A_{\bar z}=\bar\partial f-i\bar\partial\widehat h-\frac{u_R}{2\tau_2},
\end{equation}
where we treat $u_L,u_R$ as complex coefficients and define: 
\begin{equation}
u_L=\bar u,\qquad
u_R=\frac{u}{\lambda},
\qquad u\in\mathbb C,
\label{cbhcycle}
\end{equation}
which reduces to the neutral choice when $\l=1$. Following the same procedure we used for the black string, we integrate over the periodic fields and divide out the gauge volume, after which we obtain the coset partition function (see Appendix \ref{appenblackhole}): 
\begin{equation}
\label{cbhkernel}
\begin{aligned}
Z_{cs}(\tau) ={}&
\frac{R\sqrt{k\kappa}}{\mu\sqrt{\tau_2}\,|\eta(\tau)|^2}\sum_{\omega,m_X\in\mathbb Z}\int_{\mathbb C}\frac{d^2u}{\tau_2}\,
\mathcal D_{\tau}(u)\\[-1mm]&\quad\times
\exp\!\left[-\frac{\pi}{\tau_2}\left(\frac{k}{\mu}|u|^2+R^2|W|^2+\frac{2i\sqrt{k}\,\xi R}{\mu}u\bar W\right)\right].
\end{aligned}
\end{equation}
where we retain the winding in the $X$ direction as (\ref{tor:harmonics}).

The worldsheet analysis proceeds almost identically to the black string, and we use the same contour description to extrapolate the characters. We retain the same labels: $\nu$ denotes the holonomy strip, $a,\bar a$ denote the character charges and $L=a-\bar a$. We also use the integers $n,\omega $ to denote the compact boson momentum and winding. The pole structure is now controlled by: 
\begin{equation}
   B_\xi=1+2\bar a+k\nu+\mu L+i\sqrt{k}\,\xi\left(\frac nR+R\omega\right).
\label{cbhradial-pole}
\end{equation}
After shifting the contour, the crossed poles that give the discrete series yield the matching $B_\xi=1-2j$, where 
\begin{equation}
j=-\bar a-\frac{k\nu+\mu L}{2}-\frac{i\sqrt{k}\,\xi}{2}\left(\frac nR+R\omega\right), \qquad \frac12<\operatorname{Re}j<\frac{k-1}{2}.
\label{cbhdiscrete-spin}
\end{equation}
The eigenvalues of $J^3_0,\bar J^3_0$ are given by: 
\begin{align}
m&=\frac12\left[(2-\mu)L-k\nu-i\sqrt{k}\,\xi\left(\frac nR+R\omega\right)\right],
\notag\\
\bar m&=\frac12\left[-\mu L-k\nu-i\sqrt{k}\,\xi\left(\frac nR+R\omega\right)
\right]
\label{cbhcurrent-charges}
\end{align}
These are nothing but $m=j+a$ and $\bar m=j+\bar a$. As before, we define the boson momentum as: 
\begin{equation}
P_L=\frac nR+R\omega+\frac{2i\xi L}{\sqrt{k}},
\qquad
P_R=\frac nR-R\omega
\label{cbhboson-momenta}
\end{equation}
and the coset weights can be written as: 
\begin{align}
h_{\mathrm{cs}}&=-\frac{j(j-1)}{k-2}+\frac{m^2}{k}+\frac{P_L^2}{4},
\notag\\ \bar h_{\mathrm{cs}}&=-\frac{j(j-1)}{k-2}+\frac{\bar m^2}{k}+\frac{P_R^2}{4}.
\label{cbhdiscrete-weights}
\end{align}
Including the oscillator expansions, the internal CFT and ghost factor, we have: 
\begin{equation}
h_{\mathrm{tot}}=h_{\mathrm{cs}}+N+h_i,
\qquad
\bar h_{\mathrm{tot}}=\bar h_{\mathrm{cs}}+\bar N+\bar h_i
\label{cbhtotal-weights}
\end{equation}
where the internal CFT weights are $(h_i,\bar h_i)$. Similar computation for the continuous series representation can be performed, and is provided in appendix \ref{appenblackhole}. Here, we will only focus on the discrete series, which contains the candidate gravitational sector. 

\subsection{Two towers of states}
In this section, we identify two neutral towers of states in the discrete series whose contribution develops a logarithm near extremality. First we restrict our attention to the discrete series representation, which is likely to contain the Schwarzian contribution \cite{Ferko:2024uxi,Toni:2026brc}. To identify the gravitational contributions, we first restrict to the sector with vanishing compact boson momentum and winding $n=\omega=0$, and take the internal weights to be $h_i=\bar h_i=0$. This simplifies the expression for the pole parameter and the spin $j$: 
\begin{equation}
B_\xi=1+2\bar a+k\nu+\mu L,
\qquad
j=\frac{1-B_\xi}{2}
=-\bar a-\frac{k\nu+\mu L}{2}.
\label{cbhpole}
\end{equation}
together with $\frac12<j<\frac{k-1}{2}$. The total conformal weights are now given by: 
\begin{align}
    h_{\mathrm{tot}}&=-\frac{j(j-1)}{\kappa}+N+\frac{\mu^2L^2}{4k}
-\frac{(2-\mu)\nu L}{2}+\frac{k\nu^2}{4} \notag \\
 \bar h_{\mathrm{tot}}&=-\frac{j(j-1)}{\kappa}+\bar N+\frac{\mu^2L^2}{4k}
+\frac{\mu\nu L}{2}+\frac{k\nu^2}{4}
\end{align}
In the extremal limit, $\mu\to0$, and we need to identify states that become marginal in this limit, i.e. their conformal weights go to one. Since the metric perturbation is given by $\delta G_{MN}(X)\,\partial X^M\bar\partial X^N$, this leads us to consider states with $N=\bar N=1$. Level matching then requires $\nu=0$ for non-zero $L$. In the extremal limit, this gives: 
\begin{align}
     h_{\mathrm{tot}}=\bar h_{\mathrm{tot}}=-\frac{j(j-1)}{\kappa}+1=1
\end{align}
This can only be achieved by $j=1$, since $j=0$ is outside the discrete window. From (\ref{cbhpole}), this condition imposes $\bar a=-1$ in the extremal limit. Note that to have a non-vanishing state we need $a\geq-N$, which in this case gives $L=a+1\geq0$. Since we have restricted to the states with $L\neq 0$, this tower of states has: 
\begin{align}
  L>0,\qquad  \p{N,\bar N, a, \bar a,\nu}=\p{1,1,L-1,-1,0}
\end{align}
Note that the above states only contain positive values of $L$, and we must also account for the states with $L<0$, since the original sum is over $L\in \mathbb Z$. To correctly obtain the $L<0$ states, we need to flip both $a,\bar a\to -a,-\bar a$, which gives: 
\begin{align}
    a'=1-L, \quad \bar a'=+1
\end{align}
This also changes the oscillator levels to $N'=N+a $ and $ \bar N'=\bar N+\bar a$. Since the original character is expanded in $\mathcal D^+$, we can use $\widehat{\mathcal D}^{+,\nu-1}_{j}\simeq \widehat{\mathcal D}^{-,\nu}_{k/2-j}$ to write: 
\begin{align}
    j'=\frac{k}{2}-j, \quad \nu'=\nu-1
\end{align}
This yields the $L<0$ family: 
\begin{align}
    \qquad  \p{N,\bar N, a, \bar a,\nu}=\p{-L,0,L+1,1,-1}
\end{align}
and the two spins are simply given by: 
\begin{align}
    j_+=1-\frac{\mu|L|}{2}, \qquad j_-=\frac{k}{2}-1+\frac{\mu|L|}{2}
\end{align}
Now that we have all the ingredients, we can move on to count the states. As before, we can expand the analytic torsion using (\ref{torsionexpand}), and combining with the expansion of the remaining ghost contribution yields the net tower multiplicity: 
\begin{align}
   d_{\mathrm{tower}}(L)   =[q]\!\left[P(q)S_{|L|-1}(q)\right] =
\begin{cases}
0,&|L|=1\\
1,&|L|\ge2
\end{cases}
\end{align}
where we see that it is one for all $\abs{L}\geq2$, while the coefficient of $q$ vanishes for $\abs{L}=1$. Thus, the $\abs{L}=1$ mode is absent in the expansion, which is very similar to the tower of vacuum Virasoro descendants $L_{-n}\ket{0}$ with $n\geq2$, making the product of the vacuum character start at $n=2$, and similarly the omission of the $n=0,\pm1$ modes in the Schwarzian theory, where they are removed by the $SL(2,\mathbb R)$ gauge redundancy. This has been shown to play a crucial role in the exact coefficient of the $\frac{3}{2}\log T$ for the near extremal black hole entropy
\cite{Ghosh:2019rcj,Iliesiu:2020qvm,Kapec:2023ruw,Rakic:2023vhv}, and here we observe a similar mechanism for the two dimensional charged black hole, which turns out to play an important role in extrapolating the expected coefficient. It is thus tempting to identify these states as candidate gravitational towers. Note that the discrete spin window then imposes an upper bound on $\abs{L}$: 
\begin{align}
      2\le |L|<\frac1\mu
\end{align}
and the total weights can be written as: 
\begin{align}
h_{\mathrm{tot}}=\bar h_{\mathrm{tot}}=1+\frac{\mu|L|(k-\mu|L|)}{2k\kappa}=1+\frac{\delta}{2}
\end{align}
Using the above weights, we find our towers of states are contained in the full amplitude, and have the following contributions: 
\begin{align}
   \mathcal A_1 \supset \mathcal{A}^{\mathrm{disc}}=2\sum_{L=2}^{M(\mu)}I(\delta_{|L|}), \quad M(\mu)\sim\frac{1}{\mu}
\end{align}
where the factor of two comes from accounting for both signs of $L$, and we defined the integral as: 
\begin{align}
    I(\delta)=\frac12\int_{\mathcal F}\frac{d^2\tau}{\tau_2}\, e^{-2\pi\tau_2\delta}
\end{align}
To perform the modular integral, we split the fundamental domain into the bounded part $\t_2<1$ and the infinite strip $\t_2\geq1$, and the logarithm comes from the cusp in the large $\t_2$ region. Here we will only present the results and the detailed computation of the integral and the remaining $L$ sum can be found in Appendix \ref{appentower}. Define $\upsilon=\mu L$. Performing the integral yields:
\begin{equation}
     2I\!\left(\frac{\upsilon(k-\upsilon)}{k\kappa}\right)
    =  -\log\upsilon+\mathcal R_k(\upsilon),
\end{equation}
where $\mathcal R_k$ is a smooth function whose explicit form can be found in (\ref{cbh:remainder11}), and it does not contribute to the logarithm. Performing the sum over $L$ thus gives: 
\begin{align}
    -\sum_{L=2}^{M(\mu)}\log(\mu L)=\frac1\mu+\frac32\log\mu+O(1)
\end{align}
where we have expanded near extremality. Note that it is important that the sum starts from $L=2$; had we included the $L=1$ contribution, the coefficient would become $1/2$, just as in the case of the Virasoro vacuum character if we started the sum from $L_{-1}$ where one gets a $\log T^{1/2}$. The temperature of the two dimensional black hole is given by $T=\mu/(2\pi \sqrt{k})$ \cite{Giveon:2004rw},  and restoring the temperature explicitly yields: 
\begin{align}
    \mathcal{A}^{\mathrm{disc}}=\frac{\mathcal C_k}{2\pi \sqrt{k}T}+\frac{3}{2}\log\p{\frac{T}{T_0}}+\mathcal{O}(1)
\end{align}
where $\mathcal C_k$ is independent of temperature at fixed $k$, and $T_0$ is the reference scale. Note that at fixed $L$, the deviation from the extremality for the weight can be expanded as: 
\begin{align}
    \d_L=\frac{2\pi \sqrt{k}}{\kappa}\abs{L}T+\mathcal{O}(T^2)
\end{align}
which is very similar to the lifting of the zero modes of the extremal throat, where the eigenvalue at small temperature yields $nT$ for $n\geq2$ \cite{Kapec:2023ruw,Rakic:2023vhv}. However, to establish these modes as the large diffeomorphism of the throat will require us to construct the corresponding spacetime spectrum and vertex operators, which we leave for future work. 

Two related comments are in order. First, we did not include the sector $L=0$ in the spectrum, which has $j=1$ and $h_{\mathrm{tot}}=\bar h_{\mathrm{tot}}=1$ for all $\mu$. The cusp contribution is a temperature independent logarithmic divergence which does not affect the $\log T$ coefficient. In the Schwarzian sector the $n=0$ mode is also absent, so it is natural to argue for their omission as dropping some temperature independent gauge redundancy since they stay marginal for all $\mu$. Second, we have only focused on the case when the internal CFT has $(h_i,\bar h_i)=(0,0)$. However, if the internal sector contains holomorphic currents of the form $(h_i,\bar h_i)=(1,0)$ \cite{Ferko:2024uxi}, the mass-shell condition gives $(N,\bar N)=(0,1)$. This yields the same weight $\d_L$, but the expansion now has $[q^0]\,P(q)S_{L-1}(q)=1$ for $L\geq 1$. The same is found for the $L<0$ tower. This means that the sum now starts from $\abs{L}=1$, and resembles the modes $J_{-n},\,n\geq1$ of the $U(1)$ current algebra: 
\begin{align}
    -\sum_{L=1}^{M(\mu)}\log(\mu L)=\frac1\mu+\frac12\log\mu+O(1)
\end{align}
This indeed matches the contribution expected from the $U(1)$ gauge modes in the one-loop determinant around a saddle with fixed asymptotic
chemical potential \cite{Kolanowski:2024zrq}. Thus, the above coefficient $3/2$ should be understood as the gravitational contribution, while the total coefficient will depend on the internal CFT/gauge content, as well as on what charges ought to be fixed \cite{Iliesiu:2020qvm}.

Although we have found two towers of states that produce the expected logarithmic correction, we have not established it as the correction to the complete black hole entropy \cite{Giveon:2004zz}. This difficulty was already identified in \cite{Toni:2026brc}. If we wish to continue the full partition function to extremality, we need to take into account all the states. For real $\xi$, the Gaussian in \eqref{cbhkernel}  can be written as:
 \begin{align}
&\operatorname{Re}\left[
    \frac{k}{\mu}|u|^2+R^2|W|^2
    +\frac{2i\sqrt{k}\,\xi R}{\mu}u\bar W\right] = \frac{k}{\mu}
    \left|u-\frac{i\xi R}{\sqrt{k}}W\right|^2 +\frac{R^2(2\mu-1)}{\mu}|W|^2.
\label{cbhconvergence}
\end{align}
Note that the holonomy integral and the winding sum converge absolutely only if $2\mu-1>0$. In terms of temperature this gives the bound: 
\begin{equation}
    T>T_*,\qquad T_*=\frac{1}{4\pi\sqrt{k}}.
\end{equation}
which agrees with the bound found in \cite{Toni:2026brc} at leading order $1/k$. Thus, we cannot simply take the extremal limit without specifying a continuation prescription. For our selected towers of states, we can follow them directly to extremality since the discrete poles of our two towers of states only appear below $T<T_*$. This can be seen from the condition $2\leq \abs{L}<1/\mu$, which does not admit modes when $\mu\geq\frac{1}{2}$. At $\mu=\frac 12$, the first pair $L=\pm2$ sits exactly at the two endpoints of the discrete window $j_\pm$. Moreover, it can be shown that the combined contribution of the discrete poles together with the continuum continues smoothly through $T_*$, and the accompanying continuum in fact smooths the threshold crossing and only changes the non-logarithmic terms like $\mathcal{C}_k$, thus preserving the logarithmic coefficient, as explicitly checked in appendix \ref{paired-towers}. However, this does not mean such a prescription is valid for continuing the complete partition function to each character contribution before summing over all labels. Thus, to establish the logarithmic coefficient in the entropy, we must show that the remaining sectors of the full amplitude do not modify the logarithm coefficient after taking the temperature derivative to compute the entropy, and provide the corresponding continuation prescription. For now, we will only treat these two towers of states as candidates for the worldsheet origin of the expected Schwarzian correction, while identifying them with the target space Schwarzian and establishing its universality in the full near-extremal entropy are left for future work.

\subsection{Three-dimensional black string}
Now, we turn to the discussion of the three-dimensional black string and its near extremal properties. Unlike the two-dimensional charged black hole, the black string has a classical entropy
\begin{align}
     S_{\mathrm{cl}} =\frac{\pi R}{2G_3\sqrt{k}}\sqrt{M^2-Q^2}=\frac{\pi RM}{2G_3\sqrt{k+c^2}}.
\end{align}
which vanishes in the extremal limit $M=Q$. On the other hand, one can obtain a non-vanishing extremal entropy by scaling $R= c\ell$ with $\ell$ fixed in the extremal limit \cite{Kaloper:1998vw}: 
\begin{align}
    S_{\mathrm{cl}}\longrightarrow \frac{\pi M_*\ell}{2G_3}, \qquad M_*=M|_{c=\infty}
\end{align}
This however changes the asymptotic geometry in the extremal limit, since it is more natural to take the extremal limit of the HH black string with fixed $R$. The vanishing of the classical entropy is different from ordinary black holes, which naively satisfies the third law of black hole thermodynamics\footnote{For now we will treat the third law as saying the vanishing of the classical entropy in the zero temperature limit but it is not clear to us what it actually means in the quantum mechanical setting.}. However, whether the near extremal entropy contains a logarithmic correction similar to the black hole will depend on the actual quantum spectrum, rather than being determined alone by the vanishing of its classical entropy. Thus, it is motivating to search for the $\log T$ correction in the worldsheet states also for the black string geometry. 

Following the analysis of the two-dimensional charged black hole analysis, we start the search in the discrete states, and look for states that become marginal in the extremal limit (with zero internal weights). For the Horne--Horowitz geometry, the spatial radius $R$ is fixed in the extremal limit, and the proper radius of the spatial circle at the event horizon is given by: 
\begin{align}
    \tilde{R}=R\sqrt{\frac{k}{K}}=2\pi \sqrt{k} R\,T_{HH}
\end{align}
which vanishes in the extremal limit at fixed $R$. This means that the winding strings contribution to the conformal weight vanishes as $T_{HH}\to0$ ($c\to\infty$), making their weight approach marginality in the extremal limit. Thus, we should keep the winding label $\omega$. This is different from the previously discussed charged black hole, where we set the winding to zero since $X$ represents the internal compact boson after dimensional reduction. On the other hand, the momentum label $n$ grows as we approach extremality due to the term $cn/R$, and we set it to zero as for the black hole case. The metric perturbation is given by $N=\bar N=1$, and we first look in the unflowed sector with $\nu=0$, whose conformal weights are: 
\begin{align}
    h_{\rm tot}=\bar h_{\rm tot} =1-\frac{j(j-1)}{\kappa} +\frac{L^2+2icLR\omega+kR^2\omega^2}{4(k+c^2)}
\end{align}
Note that now $j$ is real, but the weights are still complex. Overcoming the initial panic, one realizes that the mass-shell condition only selects states with $L=0$. This can be seen in the extremal limit $c\to \infty$ where the mass-shell condition yields: 
\begin{align}
    1-\frac{j(j-1)}{\kappa}=1
\end{align}
which can only be achieved by $j=1$, since the discrete window discards the other solution. Using the fact that a non-vanishing state has $a,\bar a\geq-1$, together with $B=1-2j=1+a+\bar a$, the only possible solution is thus given by: 
\begin{align}
    a=\bar a=-1, \implies L=a-\bar a=0,
\end{align}
and the weights now become: 
\begin{align}
    h_{\rm tot}=\bar h_{\rm tot}=1+\frac{\tilde{R}^2\,\omega^2}{4}
\end{align}
Given that the above states are obtained in the lowest weight representation $\widehat D^+$, we can use the isomorphism $\widehat{\mathcal D}^{+,\nu-1}_{j}\simeq \widehat{\mathcal D}^{-,\nu}_{k/2-j}$ to obtain the states in the highest-weight representation: 
\begin{align}
    (\nu,a=\bar a,N=\bar N,j)=(-1,1,0,\frac{k}{2}-1)
\end{align}
which yields the same weights. Now that we have identified the candidate states, we can move on to the counting using the expansion $P(q)S_a(q)=\sum_{N=0}^{\infty}d_+(N,a)\,q^N$: 
\begin{align}
    d_{\mathrm{tower}}(\omega) = \underbrace{\bigl([q^1]\,P(q)S_{-1}(q)\bigr)^2}_{\nu=0}
+ \underbrace{\bigl([q^0]\,P(q)S_{1}(q)\bigr)^2}_{\nu=-1} =1+1=2, \quad \omega\in \mathbb Z
\end{align}
where each family contributes a single state for every winding. We can thus separate the temperature independent piece $\omega=0$ (similar to how we discarded the $L=0$) that remains marginal for any temperature, and sum over $\omega\geq1$ with a factor of four\footnote{One of the two comes from combining $\nu=0$ and $\nu=-1$, and the other one comes from summing over $\omega\geq1$ instead of $\omega\in \mathbb Z$.}: 
\begin{align}
    \mathcal A_1
\supset 4\sum_{\omega=1}^{\infty} I(\delta_\omega), \qquad
I(\delta_\omega) =\frac12\int_{\mathcal F}\frac{d^2\tau}{\tau_2}\, e^{-2\pi\tau_2\delta_\omega},
\qquad \delta_\omega=\frac{\tilde R^2\omega^2}{2}
\end{align}
Again, we can perform the integral $I(\d_\omega)$ by splitting the fundamental domain into a bounded part $\t_2<1$ and the cusp with $\t_2\geq1$. The cusp integral gives an incomplete gamma function and upon expansion near extremality yields ($\tilde{R} \to 0$ with $\omega$ fixed): 
\begin{align}
    I_{\rm cusp}(\delta_\omega)=-\log(\tilde R|\omega|) -\frac{\gamma_E+\log\pi}{2} +O(\tilde R^2\omega^2)
\end{align}
where the subleading terms are smooth functions and they admit regular Taylor expansions. Next, we turn to the sum over windings, which is no longer bounded from above. For the charged black hole, the discrete window bounds $|L|$ from above, and we could directly perform the sum. But when the accompanying continuum is introduced, $|L|$ extends to infinity and we must introduce a subtraction procedure that isolates the logarithmic correction, which is done explicitly in appendix \ref{paired-towers}. The trick lies in finding a function that yields the same logarithmic behavior in the near extremal limit, yet vanishes when the sum goes to infinity: 
\begin{align}
4\sum_{\omega=1}^{\infty}I(\delta_\omega)
={}-4\sum_{\omega=1}^{\infty}
       \log\!\left(1-e^{-\tilde R\omega}\right) +4\sum_{\omega=1}^{\infty}
  \left[I(\delta_\omega)   +\log\!\left(1-e^{-\tilde R\omega}\right)\right]
\end{align}
The first term will thus produce the logarithm while the second term is smooth near extremality and will not contribute extra logarithmic corrections. The details of the argument work identically to the charged black hole case recorded in appendix \ref{paired-towers}. Performing the sum and expanding near extremality thus yields: 
\begin{align}
    -4\sum_{\omega=1}^{\infty}\log\!\left(1-e^{-\tilde R\omega}\right)
= \frac{2\pi^2}{3\tilde R}
+2\log\tilde R -2\log(2\pi) +O(\tilde R)
\end{align}
Restoring the temperature explicitly, we obtain the contribution of these chosen states to the amplitude as: 
\begin{align}
    \mathcal A_1
\supset 4\sum_{\omega=1}^{\infty}I(\delta_\omega)
=
\frac{1}{2\pi\sqrt{k}\,R\,T_{\rm HH}}
\int_{\mathcal F}\frac{d^2\tau}{\tau_2^{3/2}}
+2\log\!\left(\frac{T_{\rm HH}}{T_0}\right)
+O(1),
\end{align}
where the integral in the first term can be performed as the incomplete beta function, and it comes from the smooth remainder that does not produce any extra logarithms. The second term gives the familiar $\log T$ correction but now with a factor of two for the black string case. We can similarly attach these discrete poles with their accompanying continuum and find that the logarithm is still preserved. Unlike the charged black hole, the full amplitude of the black string does not suffer from the convergence problem in the winding sector, and the states can be brought to extremality without any obstruction. However, to identify the coefficient as the universal correction to the full near-extremal black string entropy will require an explicit check that the remaining sectors do not contribute to the logarithm, as for the charged black hole.

It is also instructive to compare the above winding towers with the AdS$_3$ boundary graviton contributions. In particular, since the BTZ black hole and the black string are T-dual of each other, it is interesting to see how the contribution of the $\log T^{3/2}$ correction to the BTZ black hole gets transformed in the black string case. In the ungauged limit with $b$ fixed, the partition function reproduces that of thermal AdS$_3$. Following \cite{Ferko:2024uxi}, we can extrapolate the graviton sector as\footnote{This graviton sector describes the boundary gravitons of AdS$_3$ \cite{Brown:1986nw}. From the dual CFT$_2$ side, the partition function is the product of the left and right moving vacuum Virasoro character, with the vacuum-energy factors omitted. The same contribution has been independently obtained using the heat kernel method in the bulk \cite{Giombi:2008vd} (see \cite{Rakic:2025had} for an analysis of supergravity in AdS$_3$). }: 
\begin{equation}
    Z_{\mathrm{grav}}^{(x)}(b)
    =\prod_{m=2}^{\infty}(1-e^{-bm})^{-2},
    \qquad
  \log Z_{\mathrm{grav}}^{(x)}(b)
    =\frac{\pi^2}{3b}+3\log b+O(1).
\end{equation}
where the logarithm comes from taking an additional \textit{ad hoc}
$b\to 0$ limit (recall that to obtain the ungauged limit we took $R$ to infinity at fixed $b$). Thus, the contribution of the boundary graviton responsible for the near-extremal BTZ correction might formally be able to account for the black string correction as well, given $ T_{\mathrm{BTZ}}=2\pi r_+ T_{\mathrm{HH}}^2$. However, this extrapolation is quantitatively different from the winding states studied above and far less solid, since to obtain the MOS partition function we need to keep $b$ fixed while making $R\sim c\to \infty$, which will not yield a logarithm in the winding string sector above. Thus, to take this coefficient seriously one needs a justification of the \textit{ad hoc}
$b\to 0$ limit, as well as a controlled study of the complete discrete and continuous spectrum in this limit to isolate the graviton sector, and its relation to the actual HH thermal amplitude defined using the HH-time. We will leave the spacetime spectrum of the Horne-Horowitz black string for future work.

\section{Discussion}

In this paper, we studied various aspects of the Euclidean partition function of the $(SL(2,\mathbb R)\times U(1))/U(1)$ gauged WZW model and its worldsheet and spacetime spectra. Following the analysis of Hanany, Prezas and Troost \cite{Hanany:2002ev}, we recovered the expected discrete and continuous series character decomposition, as well as the finite terms that arise from retaining the full oscillator expansion. We obtained complex weights due to the choice of our gauge subgroup, which nonetheless gives a consistent spacetime spectrum. We chose the compact $X$-direction to unwind as our Lorentzian time direction, which allowed us to evaluate the modular integral and recover the regulated single string excitation trace of the horizonless geometry. The resulting energies agree with the $J_3\bar J_3$-deformed spectrum studied in \cite{Chakraborty:2024mls}, and reduce to the Maldacena--Ooguri--Son energies in the ungauged limit. One subtlety we encountered lies in the finite continuous density. Although the primary contribution reproduces the familiar gamma function density, the full oscillator expansion yields additional remainder terms that survive the ungauged limit. We believe that this might be caused by the different regularization procedures used here and in \cite{Maldacena:2000kv}. But to verify this we need to identify the boundary conditions of the regulator and match the discrete and continuous states, as well as having a consistent treatment of the discrete poles at the continuum thresholds. Hopefully this would help clarify the physical meaning of these remainder terms and the relationship between the two regulated traces. We will leave this for future work. 

Another aspect of our analysis focuses on the near extremal limit of the partition function, motivated by the near-extremal behavior of the target space geometries. For the two dimensional black hole, we identified two towers of neutral states whose weights approach one in the near extremal limit. After performing the modular integral and the corresponding sum, their combined contribution produces the expected $\log T^{3/2}$ from the Schwarzian. The $\t_2$ cusp region provides the logarithm for each near extremal mode, and the discrete spin window admits an increasing number of modes as $\mu \to 0$. Summing over them with the net multiplicities after including the ghosts gives the coefficient $3/2$. In particular, just like the Virasoro vacuum character, the absence of the $|L|=1$ modes is essential. The similarity with the Virasoro vacuum character might suggest that these towers of states describe the Schwarzian modes of the AdS$_2$ throat from the worldsheet perspective. Interestingly, although individual conformal weights of the states depend on $k$, the final logarithmic coefficient is independent of it, meaning that the computation does not require a large-$k$ limit. Moreover, towers built from a holomorphic internal current have a sum starting from $\abs{L}=1$, and contribute a term $\frac{1}{2}\log T$. This matches the single saddle contribution of the $U(1)$ gauge zero modes \cite{Iliesiu:2020qvm,Kolanowski:2024zrq}. The total coefficient thus depends on the gauge content of the internal CFT and the ensemble choices. Similarly, following the same strategy of identifying gravitational fluctuation of states that become marginal in the extremal limit, we are able to identify two winding discrete string states that contain a $\log T_{HH}^2$ contribution to the black string one-loop amplitude.

However, establishing these contributions as the correction to the complete entropy requires further work. Although we have shown that our towers with combined discrete states with the continuum terms at the thresholds do not modify the logarithm coefficient, we still need to verify that the rest of the string state will not change the logarithmic correction near extremality. We believe the logarithmic coefficient obtained from these towers will survive the sum over the rest of the states, and an explicit construction in a superstring embedding, or more explicitly in the spacetime spectrum, as done recently in \cite{Toni:2026vly} for the charged black hole, will perhaps yield a clearer picture and solidify the roles of these towers. Furthermore, for the charged black hole, we still need to specify a continuation prescription for all states when the winding sector loses absolute convergence at $T_*=\frac{1}{4\pi \sqrt{k}}$. Our selected towers only contribute to the discrete window exactly below the bound $T_*$, and they continue smoothly below this threshold with their accompanying continuum. It would be interesting to understand if this divergence bears any physical significance, and obtain a continuation prescription for the rest of the states.

\acknowledgments
I would like to thank Lorenzo Toni for introducing me to the paper \cite{Witten:1991yr} a couple of years ago, and for many interesting discussions over the years. I would also like to acknowledge AI models ChatGPT and Claude for checking some computations, searching references, and language/format editing.

\appendix

\section{Holonomy and modular integrals}
\label{intappendix}
In this appendix, we provide a detailed account of the holonomy and modular integrals used to obtain the spacetime spectrum. We first integrate over one holonomy integral and the worldsheet moduli, leaving us a finite holonomy cell. The discrete series contribution arises when the allowed interval lies in the cell, while the continuous contribution comes from the endpoints. After unfolding, the path integral is: 
\begin{align}
z_{c}={}&
R\sqrt{\kappa K}
\int_0^\infty\frac{d\tau_2}{\tau_2^{5/2}}
\int_{-1/2}^{1/2}d\tau_1\,
|\eta|^2 Z_{\mathrm{int}} \notag
\\
&\times
\int_{\mathbb C}d^2u\,\mathcal D_\tau(u)
\exp\!\left[
-\frac{\pi}{\tau_2}
\left(k|u|^2+|R+icu|^2\right)
\right],
\qquad R=\frac{bc}{2\pi},
\end{align}
As discussed in the main text, we divide the holonomy into strips: 
\begin{align}
u=(\nu+\sigma)\tau-b_0,
\qquad
\nu\in\mathbb Z,\quad 0<\sigma<1,\quad b_0\in\mathbb R.
\end{align}
The real and imaginary parts, as well as the holonomy measure, are given by : 
\begin{align}
\operatorname{Re}u=(\nu+\sigma)\tau_1-b_0, \qquad \operatorname{Im}u=(\nu+\sigma)\tau_2, \qquad d^2u=\tau_2\,d\sigma\,db_0.
\end{align}
We write the analytic torsion in the variable $U=\s \t-b_0$ and $y=e^{2\pi iU}$ within each strip, and the $\vartheta_1$ function expansion yields: 
\begin{align}
\frac{\eta(\tau)^2}{\vartheta_1(\tau,U)}&=-iq^{-1/24}y^{1/2}\mathcal F_+(q,y), \notag \\
\mathcal F_+(q,y)&=\frac1{1-y}\prod_{n=1}^{\infty}\frac{1-q^n}{(1-yq^n)(1-y^{-1}q^n)}=\sum_{N,a}d_+(N,a)q^Ny^a.
\end{align}
We also expand the internal CFT partition function as: 
\begin{align}
Z_{\mathrm{int}}
=(q\bar q)^{-c_{\mathrm{int}}/24}\sum_I d_Iq^{h_I}\bar q^{\bar h_I},\qquad c_{\mathrm{int}}=23-\frac6\kappa.
\end{align}
where the vacuum factors combine to give $(q\bar q)^{-(1+c_{\mathrm{int}})/24}=e^{\pi\tau_2\p{4-1/\kappa}}$. In these variables, the classical Gaussian is given by: 
\begin{align}
k|u|^2+|R+icu|^2={}&K\bigl[(\nu+\sigma)\tau_1-b_0\bigr]^2 +K(\nu+\sigma)^2\tau_2^2
-2cR(\nu+\sigma)\tau_2+R^2.
\end{align}
Combining the measure, classical Gaussian, oscillator expansion, and internal CFT, we obtain: 
\begin{align}
z_c(b)
={}&R\sqrt{\kappa K}\sum_{\nu,N,\bar N,a,\bar a,I}d_I\,d_+(N,a)d_+(\bar N,\bar a) \notag
\\&\times\int_0^\infty\frac{d\tau_2}{\tau_2^{3/2}}\int_{-1/2}^{1/2}d\tau_1\int_0^1d\sigma\int_{\mathbb R}db_0
\notag \\&\times\exp\!\left[-\frac{\pi R^2}{\tau_2}+2\pi cR(\nu+\sigma)-\frac{\pi K}{\tau_2}\bigl[b_0-(\nu+\sigma)\tau_1\bigr]^2\right]
\notag \\ &\times\exp\!\left[-\pi\tau_2\left(K(\nu+\sigma)^2-2\sigma^2+2\sigma(1+a+\bar a)+2\mathcal N-A_k\right)\right]
\notag \\&\times\exp\!\left[2\pi i\tau_1(\Delta+\sigma L)-2\pi ib_0L\right]
\end{align}
where we recall the definitions
\begin{align}
\mathcal N=N+\bar N+h_I+\bar h_I,
\qquad \Delta=N+h_I-\bar N-\bar h_I,\qquad L=a-\bar a
\end{align}
Let's start with the integral over $b_0$, where the relevant terms are given by: 
\begin{align}
&\int_{\mathbb R}db_0\, \exp\!\left[
-\frac{\pi K}{\tau_2}\bigl(b_0-(\nu+\sigma)\tau_1\bigr)^2 -2\pi ib_0L \right] \notag \\ &\qquad= \sqrt{\frac{\tau_2}{K}}\, e^{-\pi\tau_2L^2/K}
e^{-2\pi i(\nu+\sigma)\tau_1L}
\end{align}
The phase cancels the $\s$ dependent part of the $\t_1$ phase, so that:
\begin{align}
\int_{-1/2}^{1/2}d\tau_1\,e^{2\pi i\tau_1(\Delta-\nu L)}=\delta_{\Delta,\nu L}
\end{align}
and the $\t_1$ integral simply imposes the level matching condition. The partition function now reduces to: 
\begin{align}
z_c(b)
={}&R\sqrt{\kappa} \sum_{\nu,N,\bar N,a,\bar a,I}
d_I\,d_+(N,a)d_+(\bar N,\bar a)\,
\delta_{\Delta,\nu L} \notag
\\ &\times \int_0^1d\sigma\,
e^{2\pi cR(\nu+\sigma)} \int_0^\infty\frac{d\tau_2}{\tau_2}\,
\exp\!\left[-\frac{\pi R^2}{\tau_2}-\pi\tau_2\mathcal Q_{c,\nu}(\sigma)\right],
\end{align}
where we have defined
\begin{align}
\mathcal Q_{c,\nu}(\sigma) &=K(\nu+\sigma)^2-2\sigma^2 +2\sigma(1+a+\bar a)+2\mathcal N-A_k+\frac{L^2}{K} \notag \\
A_k&=4-\frac{1}{\kappa}
\end{align}
Next, we perform the integral over $\tau_2$. Collecting the relevant terms: 
\begin{align}
    I(\mathcal Q) =\int_0^\infty\frac{d\tau_2}{\tau_2}
    \exp\!\left[-\frac{\pi R^2}{\tau_2}-\pi\mathcal Q(\s)\tau_2\right].
\end{align}
For $\mathcal Q(\s)>0$, the above integral is simply a modified Bessel function. For $\mathcal Q(\s)<0$, it produces a tachyonic infrared growth. We work in the convergent region and obtain: 
\begin{align}
 I(\mathcal Q)=2K_0\!\left(2\pi R\sqrt{\mathcal Q}\right).
\end{align}
Using $b=2\pi R/c$, we can make the thermal parameter explicit, and write for each cell: 
\begin{align}
    z_{\mathrm{cell}}(b)=\frac{bc\sqrt{\kappa}}{\pi}\int_0^1d\sigma\,
e^{bc^2(\nu+\sigma)}
K_0\!\left(bc\sqrt{\mathcal Q_{c,\nu}(\sigma)}\right)
\end{align}
To extrapolate the spectrum, we use the integral representation for the Bessel function: 
\begin{align}
z_{\mathrm{cell}}(b)
={}&\frac{bc\sqrt{\kappa}}{\pi}
\int_0^1d\sigma
\int_{c\sqrt{\mathcal Q(\sigma)}}^\infty
d\varepsilon \hspace{1mm}
\frac{e^{-b[\varepsilon-c^2(\nu+\sigma)]}}
{\sqrt{\varepsilon^2-c^2\mathcal Q_{c,\nu}(\sigma)}}.
\end{align}
Note that the exponential becomes $-bE$ if we define $E=\varepsilon-c^2(\nu+\sigma)$ and the integral becomes: 
\begin{align}
z_{\mathrm{cell}}(b)&=b\int_{E_0}^{\infty}dE\,e^{-bE}\,
\mathscr N_{\mathrm{cell}}(E), \notag \\
\mathscr N_{\mathrm{cell}}(E)&=\frac{\sqrt{\kappa}}{\pi}
\int_{\substack{0\leq\sigma\leq1\\E+c^2(\nu+\sigma)\geq
c\sqrt{\mathcal Q_{c,\nu}(\sigma)}}}\frac{d\sigma}{\sqrt{\frac{[E+c^2(\nu+\sigma)]^2}{c^2}-\mathcal Q_{c,\nu}(\sigma)}}
\end{align}
Here the lower limit is the minimum energy of the cell: 
\begin{align}
E_0=E_{\min}^{\mathrm{cell}} =\min_{0\leq\sigma\leq1}\left[c\sqrt{\mathcal Q_{c,\nu}(\sigma)}-c^2(\nu+\sigma)\right]
\end{align}
Integration by parts on the $E$ integral separates the threshold contribution and the energy-dependent part: 
\begin{align}
\label{intbypart}
b\int_{E_0}^{\infty}dE\,e^{-bE}\mathscr N_{\mathrm{cell}}(E)
={}&e^{-bE_0}\mathscr N_{\mathrm{cell}}(E_0)+\int_{(E_0,\infty)}
e^{-bE}\,d\mathscr N_{\mathrm{cell}}(E)
\end{align}
The general structure of discrete and continuous representations is clear. If the minimum occurs inside the cell, it produces the discrete contribution. Once the minimum reaches the end points of the cell it will start to vary continuously with the energy.  

\subsection{Discrete representation}
First let's assume that the minimum $E_0$ occurs inside the range $0<\s_*<1$. Using $dE/d\s=0$ and writing $d=1+a+\bar a+k\nu$ together with (\ref{definitionofA}), we have: 
\begin{align}
    \frac{dE_0}{d\s}=\frac{c\p{(c^2+\kappa)\sigma+c^2\nu+d}}{\sqrt{\mathcal Q}}-c^2=0 \implies (c^2+\kappa)\sigma_*+c^2\nu+d=c\sqrt{\mathcal Q(\s_*)}
\end{align}
Using $c\sqrt{\mathcal Q (\sigma_*)}=E_0+c^2(\nu+\sigma_*)$, this becomes: 
\begin{align}
E_0=d+\kappa\sigma_* \implies \sigma_*=\frac{E_0-d}{\kappa}.
\end{align}
Identifying $E_0$ with the discrete energy will produce the expected discrete spin window. This can be explicitly seen by using the discrete charge relation:
\begin{align}
    E_{\mathrm{disc}}=2j+k\nu+a+\bar a
\end{align}
Setting $E_0=E_{\mathrm{disc}}$ gives $E_0=2j-1+d$, and the requirement that the minimum lies in the $\s$-cell becomes: 
\begin{align}
0<\sigma_*<1 \quad \implies \quad \frac12<j<\frac{k-1}{2}.
\end{align}
At $E=E_0$, the denominator of the integral vanishes at $\s=\s_*$ : 
\begin{align}
    \frac{[E_0+c^2(\nu+\sigma_*)]^2}{c^2} -\mathcal Q_{c,\nu}(\sigma_*)=0
\end{align}
Plugging in $\s_*=\frac{E_0-d}{\kappa}$ and $E_0-d=2j-1$, we find:
\begin{align}
   \frac{E_0^2}{c^2}+2\nu E_0=k\nu^2+2\mathcal N-4-\frac{4j(j-1)}{\kappa}+\frac{L^2}{K}.
\end{align}
which is precisely the discrete energy we obtained from continuing the worldsheet result. It remains for us to perform the integral over $\s$. Denote the two roots of the denominator by $\s_\pm$. The $\mathscr N$ integral becomes: 
\begin{align*}
    \mathscr N_{\mathrm{cell}}(E)&=\frac{\sqrt{\kappa}}{\pi}\int_{\sigma_-}^{\sigma_+}\frac{d\sigma}{\sqrt{\kappa(\sigma-\sigma_-)(\sigma_+-\sigma)}}\\&=\frac1\pi \int_{\sigma_-}^{\sigma_+}\frac{d\sigma} {\sqrt{(\sigma-\sigma_-)(\sigma_+-\sigma)}}\notag \\
    &=1
\end{align*}
Thus, the first term of (\ref{intbypart}) simply becomes the expected discrete contribution $e^{-bE_{\mathrm{dis}}}$. For the discrete representations, the roots $\s_\pm$ are contained within the holonomy cell $0<\s<1$. Next, we will see what happens when a root reaches an endpoint.

\subsection{Continuous representation}
Now the two roots approach the boundary of the cell and let's use $e=0,1$ to denote the two end points. The energy at which a root reaches an endpoint is given by: 
\begin{align}
E_{\mathrm{th},e}=c\sqrt{\mathcal Q_{c,\nu}(e)}-c^2(\nu+e)
\end{align}
If the energy goes above this threshold, meaning that $\s_-<0$ or $\s_+>1$, the boundaries of the cell are no longer roots. We can introduce a continuous parameter $s\geq0$: 
\begin{align}
    \frac{[E+c^2(\nu+e)]^2}{c^2}
-\mathcal Q_{c,\nu}(e)= \frac{4s^2}{\kappa}
\end{align}
where the normalization $4/\kappa$ is chosen (by hindsight) to match the continuous energy. The energy is then given by:  
\begin{align}
E_e(s)=c\sqrt{\mathcal Q_{c,\nu}(e)+\frac{4s^2}{\kappa}} -c^2(\nu+e)
\end{align}
At the lower point, we have $e=0$ and this gives (throughout the section we will use $E_0$ to denote $E_{e=0}$): 
\begin{align} 
    E_{c,\nu}(s)
=c\sqrt{ K\nu^2+2\mathcal N-A_k
+\frac{L^2}{K}+\frac{4s^2}{\kappa}}-c^2\nu
\end{align}
where we have used $\mathcal Q(0)=K\nu^2+2\mathcal N-A_k+\frac{L^2}{K}$. This agrees with the continuous energy we obtained from the worldsheet. Similarly, at the upper endpoint $e=1$, the energy gives the spectrally flowed sector with: 
\begin{align}
\label{spectralflow1}
    \nu'=\nu+1,\quad N'=N+a,\quad \bar N'=\bar N+\bar a, \quad a'=a-1, \quad \bar a'=\bar a-1 
\end{align}
In order to compute the endpoint contribution, we first take the lower root $\s_-<0$ while the upper root $\s_+<1$ remains in the cell. The cell integral can now be written as: 
\begin{align*}
  \pi  \mathscr N_{\mathrm{cell}} &=\int_{0}^{\s_+}\frac{d\s}{\sqrt{(\s-\s_-)(\s_+-\s)}}\\
  &=\int_{\s_-}^{\s_+}\frac{d\s}{\sqrt{(\s-\s_-)(\s_+-\s)}} -\int^{0}_{\s_-}\frac{d\s}{\sqrt{(\s-\s_-)(\s_+-\s)}}\\
  &=\pi-\int^{0}_{\s_-}\frac{d\s}{\sqrt{(\s-\s_-)(\s_+-\s)}}
\end{align*}
The roots $\s_\pm$ are determined by: 
\begin{align}
    \kappa(\s-\s_-)(\s_+-\s)=\frac{(E+c^2(\nu+\s))^2}{c^2}-\mathcal Q(\s)
\end{align}
so their sum and difference are given by: 
\begin{align}
    \sigma_-+\sigma_+=\frac{2(E-d)}{\kappa}
\end{align}
and 
\begin{align}
    \sigma_+-\sigma_-=\frac{2}{\kappa}
\sqrt{(E-d)^2+\kappa\left[\frac{(E+c^2\nu)^2}{c^2}-\mathcal Q_{c,\nu}(0)\right]}
\end{align}
The subtracted integral can then be evaluated as: 
\begin{align}
  \mathcal F_0= \frac{1}{\pi} \int^{0}_{\s_-}\frac{d\s}{\sqrt{(\s-\s_-)(\s_+-\s)}}=  \frac12-\frac1\pi\arcsin \left[\frac{E-d}{\sqrt{(E-d)^2+\kappa\left[\frac{(E+c^2\nu)^2}{c^2}-\mathcal Q_{c,\nu}(0)\right]}}\right]
\end{align}
Using $E=E_{e=0}(s)$, its contribution to the continuous density can be found as: 
\begin{align}
    z^{\mathrm{cont}}_{\mathrm{cell}}(b)
=\int e^{-bE}\,d\mathscr N_{\mathrm{cell}}(E)=-\int e^{-bE} \frac{d \mathcal{F}_0(s)}{ds}ds
\end{align}
Similarly, for the upper root $\s_+>1$, we can follow the above procedure to obtain: 
\begin{align}
    \mathcal F_1(E)=\frac1\pi\int_{\sigma_-}^{1}\frac{d\sigma} {\sqrt{(\sigma-\sigma_-)(\sigma_+-\sigma)}}
\end{align}
with $E=E_{e=1}(s)$. The two endpoint contributions then take the form: 
\begin{align}
\int_0^\infty ds\,\rho_1(s)e^{-bE_1(s)} -\int_0^\infty ds\,\rho_0(s)e^{-bE_0(s)}, \quad \rho_e=\frac{d \mathcal{F}_e(s)}{ds}, \quad e=0,1
\end{align}
We can now combine these two contributions by spectrally flowing the upper endpoint sector. Recall that $E_1$ is simply the neighboring spectrally flowed energy $E_{e=0}$ with (\ref{spectralflow1}), with which we use to flow $E_1$ to $E_0$: 
\begin{align}
    \widetilde\nu=\nu-1,\quad
\widetilde a =a+1, \quad \widetilde{\bar a}=\bar a+1,\quad
\widetilde N=N-a-1, \quad \widetilde{\bar N} =\bar N-\bar a-1
\end{align}
Using the above relabeling, we obtain $\widetilde E_1(s)=E_0 (s)$ and similarly for the density $\widetilde\rho_1(s)=\rho_0(s)$. The oscillator multiplicities do in fact change under the spectral flow, and the combined contribution can be written as: 
\begin{align}
    \Big[d_+(N-a-1,a+1)\, d_+(\bar N-\bar a-1,\bar a+1)-d_+(N,a)d_+(\bar N,\bar a)
\Big]\int_0^\infty ds\,\rho_0(s)e^{-bE_0(s)}
\end{align}
where the density at the lower point is given by: 
\begin{align}
      \rho_0(s)=\frac{2\left[E_0(s)-d-sE_0'(s)\right]}{\pi\left([E_{c,\nu}(s)-d]^2+4s^2\right)}.
\end{align}
Here $E'(s)=\frac{dE}{ds}$ is an ordinary derivative, and should not be confused with the spectral flow relabeling. This derivative acts on the on-shell energy and the explicit dependence on $s$. We can now sum over the oscillator charge at fixed $L$: 
\begin{align}
\label{horizonlessdensity}
    \rho_{N,\bar N}^{(J)}(s) &=\sum_{a=-J}^{J} \mathcal D_{\mathrm{total}}\,f_{x_a}^{\mathrm{on}}(s)
    ,\quad f^{\mathrm{on}}_{x_a}(s)=\frac{2(E-d_a-sE')}{\pi[(E-d_a)^2+4s^2]}
=\frac{x_a-sx_a'}{\pi(x_a^2+s^2)}, \notag\\*
    \mathcal D_{\mathrm{total}} &=p_Np_{\bar N}-p_{\bar N}d_+(N,a)-p_Nd_+(\bar N,a-L)
\end{align}
where $d_a=1+k\nu+2a-L$,\hspace{1mm} $x_a=(E-d_a)/2$, and we used $d_+(N-a-1,a+1)=p_N-d_+(N,a)$ to simplify $\mathcal{D}_{\mathrm{total}}$. We also retain the same finite cutoff prescription as our worldsheet analysis. Note that $d_+(N,a)$ vanishes for $a<-N$, and is given by $p_N$ when $a\geq N$. We can thus separate the contribution from $a>N$ and the rest by writing: 
\begin{align}
\label{hhdecomposition}
    d_+(N,a) =p_N\mathbf 1_{a\ge0} +\Big[d_+(N,a)-p_N\mathbf 1_{a\ge0}\Big]
\end{align}
and the terms in the bracket vanish except for $-N\leq a\leq N-1$. We can similarly perform the decomposition for the right moving part, and combining them gives: 
\begin{align}
    p_Np_{\bar N}
\big[1-\mathbf 1_{a\ge0}-\mathbf 1_{a\ge L}\big] =p_Np_{\bar N} \big[\mathbf 1_{a<L}-\mathbf 1_{a\ge0}\big]
\end{align}
Plugging this back into (\ref{horizonlessdensity}), we have: 
\begin{align}
    \sum_{a=-J}^{L-1} f^{\mathrm{on}}_{x_a}(s) -\sum_{a=0}^{J} f^{\mathrm{on}}_{x_a}(s) = \frac{1}{2\pi i}\frac{d}{ds}\log\!\left[ \prod_{a=-J}^{L-1}\frac{x_a+is}{x_a-is} \prod_{a=0}^{J}\frac{x_a-is}{x_a+is} \right]
\end{align}
where we can use the on-shell charge relation to write: 
\begin{align}
   x_a(s) =\frac{E(s)-k\nu+L}{2}
   -\frac12-a =m-\frac12-a \notag
\end{align}
The product can then be performed as: 
\begin{align}
    \prod_{a=-J}^{L-1}\frac{x_a+is}{x_a-is}
=&\frac{\Gamma(J+\tfrac12+m+is)}
     {\Gamma(J+\tfrac12+m-is)}
\frac{\Gamma(\tfrac12+\bar m-is)}
     {\Gamma(\tfrac12+\bar m+is)}, \notag \\
     \prod_{a=0}^{J}\frac{x_a-is}{x_a+is}
= &\frac{\Gamma(J+\tfrac32-m+is)}
     {\Gamma(J+\tfrac32-m-is)}
\frac{\Gamma(\tfrac12-m-is)}
     {\Gamma(\tfrac12-m+is)}
\end{align}
Plugging these back into the density, and taking the $J\to \infty$ limit, we obtain: 
\begin{align}
    \rho_{N,\bar N}^{(J)}(s)
={}&\frac{2p_Np_{\bar N}}{\pi}\log J+\frac{p_Np_{\bar N}}{2\pi i}\frac{d}{ds}
\log\frac{\Gamma(\tfrac12+\bar m-is)\Gamma(\tfrac12-m-is)}
          {\Gamma(\tfrac12+\bar m+is)\Gamma(\tfrac12-m+is)} \notag\\
&-p_{\bar N}\sum_{a=-N}^{N-1}
\big[d_+(N,a)-p_N\mathbf1_{a\ge0}\big]f^{\mathrm{on}}_{x_a}(s) \notag\\
&-p_N\sum_{a=L-\bar N}^{L+\bar N-1}
\big[d_+(\bar N,a-L)-p_{\bar N}\mathbf1_{a\ge L}\big]
f^{\mathrm{on}}_{x_a}(s)
+o(1).
\end{align}
where the first line is the $\log$ divergence and the primary density, and we have restored the finite remainder from (\ref{hhdecomposition}). As a consistency check, at $N=\bar N=0$, the last two lines are simply zero, and we only have the primary density. However, even at the first oscillator level with $(N,\bar N)=(1,0)$, they start contributing corrections to the density, exactly as what we observed for the worldsheet computation. Thus, we have successfully extrapolated the spacetime spectrum as: 
\begin{align}
z_c&(b)=\notag \\*
&\sum_{\nu,N,\bar N,a,\bar a,I}
d_I\,d_+(N,a)d_+(\bar N,\bar a)\,
\delta_{\Delta,\nu(a-\bar a)}\,e^{-bE_{\mathrm{dis}}}+\sum_{\nu,L,N,\bar N,I}
d_I\,\delta_{\Delta,\nu L}\int_0^\infty ds\,\rho_{N,\bar N}^{J}e^{-bE_{c,\nu}(s)}
\end{align}
Note that the discrete sum is restricted to the spin window and runs over $a,\bar a$ separately, while for continuous series representation we sum over $L$.

\section{2D black hole: partition function and worldsheet spectrum}
\label{appenblackhole}
Here, we use $\phi_0,v_0,\bar v_0$ for the unshifted $H_3^+$ coordinates, and write the gauged action as \cite{Dei:2024uyx}: 
\begin{align}
    S_{\mathrm{gauged}}
=\frac1\pi\int d^2z\,\Big[&k(\partial\phi_0-A_z)
  (\bar\partial\phi_0-\lambda A_{\bar z})\notag\\
&+k\bigl( \partial\bar v_0+\bar v_0\partial\phi_0-A_z\bar v_0
\bigr)\bigl(\bar\partial v_0+v_0\bar\partial\phi_0-\lambda A_{\bar z}v_0\bigr)\notag\\*
&+\partial Y_0 \bigl(\bar\partial Y_0-\bar c A_{\bar z}\bigr)
+\frac{k(1-\lambda)}{2}A_zA_{\bar z}\Big].
\label{cbhaction}
\end{align}
Gauge anomaly cancellation requires the constraint  \cite{Quella:2002fk,Giveon:2004zz}: 
\begin{equation}
k=k\lambda^{\,2}+\bar c^{\,2}.
\end{equation}
and we can solve for $\lambda=\sqrt{1-\bar c^{\,2}/k}$, where $\lambda=1$ yields the uncharged limit. It is also useful to introduce the parameters: 
\begin{equation}
\xi=\frac{\bar c}{\sqrt{k}(1+\lambda)},
\qquad \mu =1-\xi^2
=\frac{2\lambda}{1+\lambda}.
\end{equation}
Note that the extremal limit of the black hole is now given by $\abs{\xi}\to 1$ or equivalently $\lambda \to 0$. As before, we can Hodge decompose the gauge field as (\ref{tor:hodge}), but now the gauging is asymmetric, and we write: 
\begin{equation}
A_z=\partial f+i\partial\widehat h
       -\frac{u_L}{2\tau_2}, \qquad
A_{\bar z}=\bar\partial f-i\bar\partial\widehat h-\frac{u_R}{2\tau_2},
\end{equation}
where we treat $u_L,u_R$ as complex coefficients and define: 
\begin{equation}
u_L=\bar u,\qquad
u_R=\frac{u}{\lambda},
\qquad u\in\mathbb C
\end{equation}
which reduces to the neutral choice when $\l=1$. Again, it is useful to introduce the fields: 
\begin{equation}
p=-if+\frac{i}{4\tau_2}(u_Lz+u_R\bar z),
\qquad
h=\widehat h+\frac{i}{4\tau_2}(u_Lz-u_R\bar z),
\end{equation}
where the gauge fields can be expressed as $A_z=i\partial(p+h)$ and $A_{\bar z}=i\bar\partial(p-h)$. Due to the asymmetric gauging, introduce the combination: 
\begin{equation}
p_\lambda=\frac{(1+\lambda)p+(1-\lambda)h}{2},
\qquad
h_\lambda=\frac{(1-\lambda)p+(1+\lambda)h}{2}
\label{hodge-combinations}
\end{equation}
and write the shifted coordinates as: 
\begin{equation}
\phi=\phi_0-ip_\lambda,
\qquad
V=e^{ih_\lambda}v_0,
\qquad
\bar V=e^{-ih_\lambda}\bar v_0,
\qquad
Y=Y_0-\frac{i\bar c}{2}(p-h).
\label{cbhfield-redefinitions}
\end{equation}
We retain the winding in the $X$ direction as (\ref{tor:harmonics}), and decompose the shifted fields into a periodic and a holonomy part: 
\begin{align}
\phi&=\widehat\phi+L_+,  \qquad
V=e^{-L_-}\widehat V, \qquad
\bar V=e^{L_-}\widehat{\bar V},
\notag\\
h&=\widehat h+\frac{i}{4\tau_2}
       \left(\bar uz-\frac{u\bar z}{\lambda}\right), \qquad
Y=\widehat Y+F+\frac{\bar c\,u\bar z}{4\lambda\tau_2}.
\label{cbhfield-holonomies}
\end{align}
These decompositions can then be plugged back into the gauged action, yielding a periodic part and a  harmonic part. The path integral over the periodic part of the field $\widehat V$ will again contain the analytic torsion (\ref{torsion}), and the gauge ghost fields determinant cancels the scalar oscillators. We thus have:
\begin{align}
Z_{cs}(\tau)
={}&
\frac{R\sqrt{k\kappa}}
{\mu\sqrt{\tau_2}\,|\eta(\tau)|^2}\sum_{\omega,m_X\in\mathbb Z}\int_{\mathbb C}\frac{d^2u}{\tau_2}\,\mathcal D_{\tau}(u)\notag\\*[-1mm]&\quad\times\exp\!\left[-\frac{\pi}{\tau_2} \left(\frac{k}{\mu}|u|^2+R^2|W|^2+\frac{2i\sqrt{k}\,\xi R}{\mu}
u\bar W\right)\right]
\end{align}
Next we study its worldsheet spectrum. First, we Poisson resum the $m_X$ winding to get: 
\begin{align*}
    &\sum_{m_X\in\mathbb Z}
\exp\!\left[
-\frac{\pi}{\tau_2}
\left(R^2|W|^2+\frac{2i\sqrt{k}\xi R}{\mu}u\bar W\right)
\right]\\
&\quad=
\frac{\sqrt{\tau_2}}{R}
\sum_{n\in\mathbb Z}
\exp\!\left[
-\pi\tau_2\left(\frac{n^2}{R^2}+R^2\omega^2\right)
+2\pi i n\omega\tau_1
-\frac{\pi k\xi^2}{\mu^2\tau_2}u^2
-\frac{2\pi\sqrt{k}\xi}{\mu}u
\left(\frac nR+R\omega\right)
\right].
\end{align*}
We retain the same strip decomposition: 
\begin{align}
u=(\nu+\sigma)\tau-b_0,
\qquad
\nu\in\mathbb Z,\quad 0<\sigma<1,\quad b_0\in\mathbb R.
\end{align}
The analytic torsion is periodic under $u\to u+\tau \nu$ and we define $U=u-\nu \t=\s \t-b_0$, so that $\mathcal D_\tau(u)=\mathcal D_\tau(U)$. As before, we write $u=x+i\tau_2(\nu+\sigma)$ and $x=(\nu+\sigma)\tau_1-b_0$. The measure then becomes: 
\begin{align}
    \int_{\mathbb C}\frac{d^2u}{\tau_2}=\sum_{\nu\in\mathbb Z}\int_0^1d\sigma\int_{\mathbb R}dx
\end{align}
We can then use the expansion for the analytic torsion (\ref{torsionexpand}) to obtain: 
\begin{align*}
Z_{\mathrm{cs}}(\tau)
={}&\frac{\sqrt{k\kappa}}{\mu|\eta(\tau)|^6}
\sum_{n,\omega,\nu\in\mathbb Z}
\sum_{a,\bar a\in\mathbb Z}
S_a(q)S_{\bar a}(\bar q)
\\
&\times
e^{-\pi\tau_2(n^2/R^2+R^2\omega^2)}
e^{2\pi i\tau_1(n\omega-\nu L)}
\\
&\times\int_0^1d\sigma\int_{\mathbb R}dx\,
\exp\!\Bigg[
2\pi\tau_2\bigl(\sigma^2-\sigma-\sigma(a+\bar a)\bigr)
+2\pi ixL
\\
&\hspace{15mm}
-\frac{\pi}{\tau_2}
\left(\frac{k}{\mu}|u|^2+
\frac{k\xi^2}{\mu^2}u^2\right)
-\frac{2\pi\sqrt{k}\xi}{\mu}\,
u\left(\frac nR+R\omega\right)
\Bigg],
\end{align*}
Performing the integral over $x$ yields
\begin{align}
I_x&=\int_{\mathbb R}dx\, \exp\!\Bigg[
-\frac{\pi k}{\mu^2\tau_2}x^2
-2\pi x\left(\frac{\sqrt{k}\xi}{\mu}
\left(\frac nR+R\omega\right)-iL +\frac{ik\xi^2}{\mu^2}(\nu+\sigma)\right)\Bigg] \notag \\
&=\mu\sqrt{\frac{\tau_2}{k}}\,
\exp\!\Bigg[\frac{\pi\tau_2\mu^2}{k}
\left(\frac{\sqrt{k}\xi}{\mu}\left(\frac nR+R\omega\right)
-iL +\frac{ik\xi^2}{\mu^2}(\nu+\sigma)\right)^2\Bigg].
\end{align}
Collecting the terms proportional to $\s$ then yields: 
\begin{align}
I_\sigma(B_\xi) =\sqrt{\kappa\tau_2}
\int_0^1d\sigma\, e^{-\pi\tau_2(\kappa\sigma^2+2B_\xi\sigma)}
\end{align}
where 
\begin{align}
B_\xi=1+2\bar a+k\nu+\mu L +i\sqrt{k}\xi\left(\frac nR+R\omega\right).
\end{align}
We can then introduce an auxiliary parameter $s$ to get rid of the quadratic $\s$ term, and the integral over $\s$ becomes elementary: 
\begin{align}
I_\sigma(B_\xi)
=\frac1\pi\int_{\mathbb R}ds\,e^{-4\pi\tau_2s^2/\kappa}
\left[\frac{e^{2\pi\tau_2(2is-B_\xi)}}{2is-B_\xi}
-\frac1{2is-B_\xi}\right]
\end{align}
The exponent can be written as: 
\begin{align}
-\frac{4\pi\tau_2}{\kappa}s^2+4\pi i\tau_2s-2\pi\tau_2B_\xi
=-\frac{4\pi\tau_2}{\kappa} \left(s-\frac{i\kappa}{2}\right)^2-\pi\tau_2(\kappa+2B_\xi)
\end{align}
where we see the Gaussian is located at $s-\frac{i\kappa}{2}$. To separate the discrete and continuous contributions, we shift the contour upward $\mathbb R\longrightarrow\mathbb R+\frac{i\kappa}{2}$. Note that the pole is located at $s_*=-\frac{iB_\xi}{2}$, and is crossed when: 
\begin{align}
0<\operatorname{Im}s_*<\frac{\kappa}{2} \implies 
-\kappa<\operatorname{Re}B_\xi<0
\end{align}
Using the same method as in the black string case, we can identify $B_\xi=1-2j$, which gives: 
\begin{align}
j=-\bar a-\frac{k\nu+\mu L}{2}-\frac{i\sqrt{k}\xi}{2} \left(\frac nR+R\omega\right).
\end{align}
As in the black string calculation, we define $m=j+a$ and $\bar m=j+\bar a$. Using $a-\bar a=L$, we can write:  
\begin{align*}
m&=\frac12\left[(2-\mu)L-k\nu
-i\sqrt{k}\xi\left(\frac nR+R\omega\right) \right],\\
\bar m&=\frac12\left[-\mu L-k\nu
-i\sqrt{k}\xi\left(\frac nR+R\omega\right)\right]
\end{align*}
Collecting the rest of the exponent gives: 
\begin{align*}
F_\nu
={}&
\exp\!\Bigg\{
-\pi\tau_2\Bigg[
\frac{n^2}{R^2}+R^2\omega^2
+k\nu^2-2\xi^2\nu L+\frac{\mu^2L^2}{k}\\*
&\hspace{24mm}-\xi^2\left(\frac nR+R\omega\right)^2
+\frac{2i\xi}{\sqrt{k}}(k\nu+\mu L)
\left(\frac nR+R\omega\right)\Bigg]\Bigg\}\times e^{2\pi i\tau_1(n\omega-\nu L)}.
\end{align*}
Since $F_\nu=q^{H_L}\bar q^{H_R}$, we can solve for $H_{L,R}$ as: 
\begin{align}
H_L=\frac{m^2}{k}+\frac{P_L^2}{4},
\qquad
H_R=\frac{\bar m^2}{k}+\frac{P_R^2}{4},
\end{align}
where
\begin{align}
P_L=\frac nR+R\omega+\frac{2i\xi L}{\sqrt{k}}, \qquad P_R=\frac nR-R\omega
\end{align}
Including the oscillator level and the internal CFT weights then yields (\ref{cbhtotal-weights}). The continuous representation then comes from the remaining integral, which is given by:
\begin{align}
\label{cbh:contint1}
        I_\sigma^{\mathrm{Rem}}(B_\xi)=\frac1\pi\int_{\mathbb R}ds\,
      e^{-4\pi\tau_2s^2/\kappa}\left[\frac{e^{-\pi\tau_2(\kappa+2B_\xi)}}
     {2is-B_\xi-\kappa}-\frac1{2is-B_\xi}\right]
\end{align}
Using the same expansion as before, and including the vacuum factors, we have: 
\begin{align}
-\frac18+\frac{s^2}{\kappa}
=
-\frac{c_{\mathrm{ws}}}{24}
+\frac{s^2+\frac14}{\kappa},
\qquad
c_{\mathrm{ws}}=\frac{3k}{\kappa}
\end{align}
where we recognize it as the spin of the continuous representation: 
\begin{align}
    -\frac{j(j-1)}{\kappa} = \frac{s^2+\frac14}{\kappa}, \qquad j=\frac12+is
\end{align}
Using $F_\nu=q^{H_L}\bar q^{H_R}$, we can solve for the continuous weights: 
\begin{align}
    h_{\mathrm{cs}}^{\mathrm{cont}} =\frac{s^2+\frac14}{\kappa} +\frac{m^2}{k}+\frac{P_L^2}{4}, \qquad \bar h_{\mathrm{cs}}^{\mathrm{cont}} = \frac{s^2+\frac14}{\kappa} +\frac{\bar m^2}{k}+\frac{P_R^2}{4}
\end{align}
To obtain the continuous density, we use the same identity to write the first term in (\ref{cbh:contint1}) as: 
\begin{align}
    \frac{F_\nu S_a(q)S_{\bar a}(\bar q)\, e^{-\pi\tau_2(\kappa+2B_\xi)}}{2is-B_\xi-\kappa}=F_{\nu+1} \frac{q^aS_a(q)\,\bar q^{\bar a}S_{\bar a}(\bar q)} {2is-B_\xi-\kappa}
\end{align}
with the following relabeling $\nu'=\nu+1, a'=a-1,\bar a'=\bar a-1$, we have: 
\begin{align}
     B_\xi' &=1+2(\bar a-1)+k(\nu+1)+\mu L +i\sqrt{k}\xi\left(\frac nR+R\omega\right) \notag \\ 
     &=B_\xi+\kappa. \notag
\end{align}
Again, using $q^{a+1}S_{a+1}(q)=1-S_a(q)$, the combined character numerator is: 
\begin{align}
    [1-S_a(q)][1-S_{a-L}(\bar q)]-S_a(q)S_{a-L}(\bar q)=1-S_a(q)-S_{a-L}(\bar q)
\end{align}
and the partition function for the continuous part is: 
\begin{align}
    Z_{cs}^{\mathrm{cont}}(\tau)={}& \frac1\pi \sum_{\nu,L,n,\omega\in\mathbb Z} \frac{F_\nu}{|\eta(\tau)|^6} \int_{\mathbb R}ds\, e^{-4\pi\tau_2s^2/\kappa} \sum_{a\in\mathbb Z} \frac{1-S_a(q)-S_{a-L}(\bar q)} {2is-B_\xi}
\end{align}
Recall we also have 
\begin{align}
    \alpha=\frac12-m,\qquad \bar\alpha=\frac12-\bar m=\alpha+L
\end{align}
with $B_\xi=2(\a+a)$. Now, we decompose the integral $s\in \mathbb R$ as $s$ and $-s$ and write:
\begin{align}
     \frac1\pi \left[ \frac1{2is-B_\xi} +\frac1{-2is-B_\xi} \right]=-f_{\alpha+a}(s)
\end{align}
where $f_x=x/\p{\pi(x^2+s^2)}$ as before. The continuous density can be written as:
\begin{align}
\label{densitysum}
    \rho_{\mathrm{ws},J}(s;q,\bar q) = \sum_{a=-J}^{J} \bigl[S_a(q)+S_{a-L}(\bar q)-1\bigr] f_{\alpha+a}(s)
\end{align}
where the primary density is given by: 
\begin{align}
  \rho_{0,J}(s)=\sum_{a=-J}^{J}
   \bigl[\mathbf1_{a\ge0}+\mathbf1_{a\ge L}-1\bigr] f_{\alpha+a}(s)
\end{align}
To perform this sum, we use $\mathbf1_{a\geq L}-1=-\mathbf1_{a\leq L-1}$ and write the sum as: 
\begin{align}
\rho_{0,J}(s)
=\sum_{a=0}^{J}f_{\alpha+a}(s)+
\sum_{b=0}^{J+L-1}f_{1-\bar\alpha+b}(s).
\end{align}
where for the second term, we have defined $b=L-1-a$ and used the fact that $f_{-x}=-f_x$. Performing the sum and taking the $J\to \infty$ limit yield: 
\begin{align}
    \rho_{0,J}(s) =\frac{2\log J}{\pi}+\frac1{2\pi i}\partial_s\log \frac{\Gamma(\alpha-is)\Gamma(1-\bar\alpha-is)}{\Gamma(\alpha+is)\Gamma(1-\bar\alpha+is)}+o(1)
\end{align}
The finite remainder is then given by the rest of the sum: 
\begin{align}
    \mathcal R(s;q,\bar q) =\sum_{a\in\mathbb Z} \bigl[S_a(q)-\mathbf1_{a\ge0}\bigr] f_{\alpha+a}(s)+ \sum_{a\in\mathbb Z} \bigl[S_a(\bar q)-\mathbf1_{a\ge0}\bigr] f_{\bar\alpha+a}(s)
\end{align}

\subsection{Two towers of states and the logarithm}
\label{appentower}
In this appendix we record several computations in detail. First, we elaborate on the state counting procedure. As before, the expansion for the analytic torsion is: 
\begin{align}
\mathcal D_\tau(U)
=
\frac{e^{2\pi\tau_2(\sigma^2-\sigma)}}{|\eta(\tau)|^4}
\sum_{a,\bar a\in\mathbb Z}
S_a(q)S_{\bar a}(\bar q)\times e^{-2\pi\tau_2\sigma(a+\bar a)}e^{2\pi i(x-\nu\tau_1)(a-\bar a)}
\end{align}
where $S_a(q)=\sum_{b\ge0}(-1)^bq^{b(b+2a+1)/2}$. Let's use $P(q)=\prod_{m\ge1}(1-q^m)^{-1}$ as the $1/\eta(\t)$ expansion stripped away the vacuum factor. Combining with the remaining boson and ghost expansion, the holomorphic part of the expansion is simply given by: 
\begin{align}
    P(q)S_a(q)=\prod_{m\ge1}(1-q^m)^{-1}\sum_{b\ge0}(-1)^bq^{b(b+2a+1)/2}
\end{align}
For the $L>0$ branch, we have $a=L-1$. When $L=1$, we can expand the above multiplicity and find that the linear terms in $q$ cancel out, and for $L\geq2$, the coefficient of $q$ is simply one. The anti-holomorphic part has $\bar a=-1$, and it does have a non-vanishing $\bar q$ term at $L=1$. Thus, multiplying the two gives the net-contribution of the $L>0$ branch, which yields a net multiplicity: 
\begin{align}
    \begin{cases}
0,&L=1,\\
1,&L\ge2.
\end{cases}
\end{align}
Similarly, the negative $L$ branch has $a=1-\abs{L}$ and $\bar a=1$. Plugging into the expansions above we again find the net multiplicity as: 
\begin{align}
\begin{cases}
0,&L=-1,\\
1,&L=-2,-3,-4,\ldots,
\end{cases}
\end{align}
The $\abs{L}=1$ mode is then absent in the expansion. The discrete spin window then imposes an upper bound on $\abs{L}$: 
\begin{align}
      2\le |L|<\frac1\mu
\end{align}
and the total weights can be written as: 
\begin{align}
h_{\mathrm{tot}}=\bar h_{\mathrm{tot}}=1+\frac{\mu|L|(k-\mu|L|)}{2k\kappa}=1+\frac{\delta}{2}
\end{align}
We then insert the weights into the one-loop amplitude $\mathcal A_1$. Our towers of states are contained in this amplitude, and have the following contributions: 
\begin{align}
    \mathcal{A}^{\mathrm{disc}}=\frac{1}{2}\sum_L\int \frac{d\t_1 d\t_2}{\t_2}\hspace{1mm}e^{-2\pi \t_2 \d}
\end{align}
The modular integral can be performed by splitting the fundamental domain into the bounded part $\t_2<1$ and the infinite strip $\t_2\geq1$. The bounded integral then has an ordinary Taylor expansion, and will not contribute to the logarithm: 
\begin{align}
    \mathcal{A}^{\mathrm{disc}}=\frac{1}{2}\sum _LI_{\mathrm{bounded}}+\frac{1}{2}\sum_L\int_1^{\infty}\frac{d\t_2}{\t_2}\hspace{1mm}e^{-2\pi \t_2 \delta}
\end{align}
The cusp integral can be performed and yields an incomplete gamma function, whose expansion near extremality yields: 
\begin{align}
    I_{\mathrm{cusp}}(\delta)
=-\frac12\log\delta-\frac12\bigl[\gamma_E+\log(2\pi)\bigr]+O(\delta).
\end{align}
Combining the $L>0$ and $L<0$ contributions simply multiplies this by a factor of two, since they have identical weights. We will take $L>0$ from now on. Recall that 
\begin{align}
    \delta=\frac{\mu L}{\kappa}\p{1-\frac{\mu L}{k}}, \quad T=\frac{\mu}{2\pi \sqrt{k}}
\end{align}
Thus, we can simply write the total integral as: 
\begin{align}
\label{appenint}
    2I(\d)=-\log \p{\mu L}+\mathcal{R}_{k}(\mu L)
\end{align}
where the second term is a smooth function of $\mu L$ and can be evaluated explicitly as: 
\begin{align}
\label{cbh:remainder11}
    \mathcal R_k(\upsilon)
=1-\gamma_E+ \log\!\left[\frac{\kappa}{3\sqrt3\,\pi(1-\upsilon/k)}\right]
+ \sum_{m=1}^{\infty} \frac{(-1)^{m+1}z^m}{m\,m!}\, B_{1/4}\!\left(\frac12,1+\frac m2\right)
\end{align}
where $z=2\pi\delta$. Note that the series converges for $0<\upsilon<1$, and thus is finite in the extremal limit. $\mathcal{R}_k(\upsilon)$ then admits an ordinary Taylor expansion: 
\begin{align}
    \mathcal R_k(\upsilon) = 1-\gamma_E+\log\!\left(\frac{k-2}{3\sqrt{3}\,\pi}\right) +
\left[ \frac{1}{k}+\frac{2\pi}{k-2}\left(\frac{\pi}{6}+\frac{\sqrt{3}}{4}\right) \right]\upsilon +\mathcal O(\upsilon^2)
\end{align}
which is analytic at $\upsilon=0$ and does not contribute any logarithm.

Now, let's focus on the logarithm term in (\ref{appenint}). Performing the sum over $L$ yields: 
\begin{align}
    -\sum_{L=2}^{M\sim\mu^{-1}}\log \p{\mu L}=-(M-1)\log\mu-\log M!
\end{align}
In the extremal limit, $M\to \infty$, and we can use Stirling's formula to approximate the second term, which gives: 
\begin{align}
    -\sum_{L=2}^{M\sim\mu^{-1}}\log \p{\mu L}=\frac{1}{\mu}+\frac{3}{2}\log \mu-\frac{1}{2}\log 2\pi +\mathcal{O}(\mu) 
\end{align}
The sum over the smooth remainder $\mathcal{R}_k(\upsilon)$ will not contribute any logarithmic terms. This can be seen by approximating the sum using the Euler–Maclaurin summation formula: 
\begin{align}
    \sum_{L=2}^{M(\mu)}\mathcal R_k(\mu L) =\frac{1}{\mu}\int_0^1\mathcal R_k(\upsilon)\,d\upsilon+O(1)
\end{align}
which contributes as $1/\mu$ without any logarithm. After restoring the temperature explicitly, the amplitude of the towers of states is now given by: 
\begin{align}
    \mathcal{A}^{\mathrm{disc}}=\frac{\mathcal C_k}{2\pi \sqrt{k}T}+\frac{3}{2}\log\p{\frac{T}{T_0}}+\mathcal{O}(1)
\end{align}
where $\mathcal C_k=1+\int_0^1\mathcal R_k(v)\,dv$ is constant in the extremal limit.

\subsection{The continuum of the two towers}
\label{paired-towers}
In the previous subsection, we have only considered the discrete contribution of the towers, which appear below the temperature $T_*$ where the full amplitude stops being convergent. However, their accompanying continuum also contributes below the threshold temperature and might change the logarithmic coefficient. Here, we will include their accompanying continuum contribution and show that it does not change the logarithm coefficient. As before, we take $k>3$ and define: 
\begin{align}
    \upsilon=\mu L, \qquad \d(\upsilon)=\frac{\upsilon(k-\upsilon)}{\kappa k}
\end{align}
To include both contributions, we go back to the original $\s$-integral before the contour deformation $I_\s(B_\xi)$. The pole parameters of the towers are given by: 
\begin{align}
    B_+=\upsilon-1, \qquad B_-=-\kappa-B_+
\end{align}
It is useful to change $\s\to 1-\s$ for the second branch, which gives the integral: 
\begin{align}
\label{identity1}
    I_\sigma(-\kappa-B)
    =e^{\pi\tau_2(\kappa+2B)}I_\sigma(B)
\end{align}
Now, including the rest of the exponential for the two towers we get for their amplitude: 
\begin{align}
    e^{-2\pi\tau_2\delta}\, e^{-\pi\tau_2 B_\pm^2/\kappa}\,I_\sigma(B_\pm)
\end{align}
and using (\ref{identity1}) gives: 
\begin{align}
    e^{-\pi\tau_2 B_-^2/\kappa}I_\sigma(B_-)
    =e^{-\pi\tau_2 B_+^2/\kappa}I_\sigma(B_+)
\end{align}
We can thus see that the two branches give the same contribution and taking both into account cancels the factor of $1/2$ in the amplitude. The combined contribution is thus: 
\begin{align}
    \mathcal I_k(\upsilon)=
    \int_{\mathcal F}\frac{d^2\tau}{\tau_2}\,
    e^{-\pi\tau_2(\upsilon^2/k+1/\kappa)}
    I_\sigma(\upsilon-1).
    \label{cbhpaired-integral}
\end{align}
where we used $2\delta(\upsilon)+B_+^2/\kappa =\upsilon^2/k+1/\kappa$. Completing the square then gives: 
\begin{align}
    \kappa\sigma^2+2(\upsilon-1)\sigma
    +\frac{\upsilon^2}{k}+\frac1\kappa =\kappa\left(\sigma-\frac{1-\upsilon}{\kappa}\right)^2+2\delta(\upsilon)
\end{align}
For small $\upsilon$, the Gaussian lies in the integration range for $\s$. If we were to extend the integration range to the entire $\mathbb R$, the contribution then yields the discrete states: 
\begin{align}
    e^{-2\pi\tau_2\delta(\upsilon)} \sqrt{\kappa\tau_2} \int_{-\infty}^{\infty}d\sigma\, e^{-\pi\kappa\tau_2(\sigma-\sigma_0)^2} = e^{-2\pi\tau_2\delta(\upsilon)}
\end{align}
Since the original integral is over $\s\in[0,1]$, from the above we can deduce that the continuous representation is in the subtracted integral (including the minus sign). By this we simply mean: 
\begin{align}
\underbrace{\int_0^1}_{\text{original interval}}
=
\underbrace{\int_{-\infty}^{\infty}}_{\text{discrete}}
-
\underbrace{\left(
\int_{-\infty}^{0}+\int_1^\infty
\right)}_{\text{continuous}}
\end{align}
Note that the continuum terms are smooth near $\upsilon\to 0^+$, and are well behaved at $\s=1,0$. Thus, the modular integrals of these terms, as well as their derivatives remain finite near zero. Thus, we can still write: 
\begin{align}
    \mathcal I_k(\upsilon)=-\log\upsilon+\text{smooth terms},
    \qquad \upsilon\to0.
\end{align}
Next we need to perform the sum over $L$, where we write: 
\begin{align}
    \mathcal A^{\mathrm{comp}}(\mu) =\sum_{L=2}^{\infty}\mathcal I_k(\mu L)
\end{align}
Note that the upper limit is now at infinity, instead of the discrete window $1/\mu$. For $\mu L>1$, the discrete series is absent while the continuum is still there. To get rid of the $\log \upsilon$ term near $\upsilon\to 0^+$, and also vanish near $\upsilon \to \infty$, we introduce the term $\log(1-e^{-\mu L})$ and write our sum as: 
\begin{align}
    \mathcal A^{\mathrm{comp}}(\mu)&= -\sum_{L=2}^{\infty}\log(1-e^{-\mu L})+\sum_{L=2}^{\infty}\mathcal H_k(\mu L), \notag \\
    \mathcal H_k(\upsilon)&=\mathcal I_k(\upsilon)+\log(1-e^{-\upsilon})
\end{align}
the function $\mathcal H_k$ is smooth at zero and decreases rapidly at infinity. The first sum can be performed and yields: 
\begin{align}
    -\sum_{L=2}^{\infty}\log(1-e^{-\mu L})
    =\frac{\pi^2}{6\mu} +\frac32\log\mu -\frac12\log(2\pi)+O(\mu).
\end{align}
after expanding $\mu\to 0$. The second term is smooth and does not produce a new logarithmic contribution. Using again Euler–Maclaurin we can write its sum as: 
\begin{align}
    \sum_{L=2}^{\infty}\mathcal H_k(\mu L) =\frac1\mu\int_0^\infty\mathcal H_k(\upsilon)\,d\upsilon -\frac32\mathcal H_k(0)+O(\mu)
\end{align}
Combining the two expressions yields:
\begin{align}
    \mathcal A^{\mathrm{comp}}(\mu)
    =\frac{\widetilde{\mathcal C}_k}{\mu}+\frac32\log\mu+\mathcal B_k+O(\mu)
\end{align}
where $\widetilde{\mathcal C}_k$ and $\mathcal B_k$ are independent of $\mu$: 
\begin{align}
    \widetilde{\mathcal C}_k =\frac{\pi^2}{6}+\int_0^\infty \mathcal H_k(\upsilon)\,d\upsilon, \qquad 
    \mathcal B_k =-\frac12\log(2\pi)-\frac32\,\mathcal H_k(0)
\end{align}
Note that $\widetilde{\mathcal C}_k$ is different from $\mathcal{C}_k$ we have obtained in the previous discrete sector computation.

\bibliographystyle{jhep}
\bibliography{refs}

\end{document}